\documentclass[fleqn,usenatbib]{mnras}

\usepackage[T1]{fontenc}

\usepackage{xcolor}
\usepackage{graphicx}
\usepackage{dcolumn}
\usepackage{bm}
\usepackage{caption}
\usepackage{subcaption}
\usepackage{tabularx}
\usepackage{amsmath}
\usepackage[normalem]{ulem}  
\usepackage{amssymb}
\usepackage{hyperref}

\usepackage{multirow}
\title[Model Independent Lensing for Sub-structures]{Model independent lensing sub-structure detection with multiply-imaged star clusters `constellations'}

\author[L. Fung et al.]{
Leo W.H. Fung,$^{1,2}$  \thanks{E-mail: leowhfung@gmail.com | wing.h.fung@durham.ac.uk}
Tom Broadhurst,$^{3,4,5}$
Sung Kei Li,$^{6,7,8}$
\newauthor
Jeremy Lim,$^{6,7}$
Giorgio Manzoni,$^{2}$
George F. Smoot$^{2,5,9,10,11}$
\thanks{Our good friend and mentor G. Smoot passed away prior to the submission of this manuscript, on 18 Sept 2025. A portion of this work was a part of L.Fung's thesis for which G. Smoot supervised.}
\\
$^{1}$Institute for Computational Cosmology \& Centre for Extragalactic Astronomy, Durham University, Stockton Rd, Durham DH1 3LE, UK\\
$^{2}$Jockey Club Institute for Advanced Study \& Department of Physics, Hong Kong University of Science and Technology, \\ Clear Water Bay, Hong Kong\\
$^{3}$Department of Theoretical Physics, University of Basque Country UPV/EHU, Bilbao, Spain\\
$^{4}$Ikerbasque, Basque Foundation for Science, Bilbao, Spain\\
$^{5}$Donostia International Physics Center, Paseo Manuel de Lardizabal, 4, San Sebasti\'an, 20018, Spain\\
$^{6}$Department of Physics, The University of Hong Kong, Pokfulam, Hong Kong\\
$^{7}$The Hong Kong Institute for Astronomy and Astrophysics, The University of Hong Kong, Pokfulam Road, Hong Kong, P. R.China\\
$^{8}$ IFCA, Instituto de F\'isica de Cantabria (UC-CSIC), Av. de Los Castros s/n, 39005  Santander, Spain\\
$^{9}$Department of Physics, University of California, Berkeley, California, USA, \it emeritus\\
$^{10}$Laboratoire APC-PCCP, Université Sorbonne Paris Cité, Université Paris Diderot, \it emeritus\\
$^{11}$Laboratoire Astroparticule et Cosmologie, Universit{\'e} de Paris, F-75013, Paris, France, \it emeritus\\
}

\date{Accepted XXX. Received YYY; in original form ZZZ}

\pubyear{\the\year{}}

\begin{document}
\label{firstpage}
\pagerange{\pageref{firstpage}--\pageref{lastpage}}
\maketitle

\begin{abstract}
A broad class of dark matter (DM) models predicts the existence of sub-structures residing in DM haloes on sub-resolvable angular scales. 
Techniques to extract such a generic feature from diffraction-limited observations are lacking. 
In this work, we propose a model-independent `super-resolving' method applicable to strong gravitational lens systems that is fully data-driven, without reference to any lens models. 
This method relies on a specific way of applying the optical Liouville theorem across multiple scales in the imaging data, a technique we refer to as geometrical-duality. 
We test this method using realistic simulations and apply it to a `constellation' consists of 11 compact star clusters seen in two giant arcs in the lensing cluster SMACS0723 imaged by JWST as a proof-of-concept. 
We find reasonable self-consistency with the expectation of non-detection given the statistical sensitivity, except for one pair of star clusters. 
Such an outlier can be explained in the context of CDM as sub-haloes lensing in the mass range $M_{\rm sub} = 10^8 - 10^9\,M_\odot$, for which the corresponding Einstein radius is smaller than the diffraction limit of JWST.
\end{abstract}

\section{Introduction}

\begin{figure*}
    \centering
    \includegraphics[width=\textwidth]{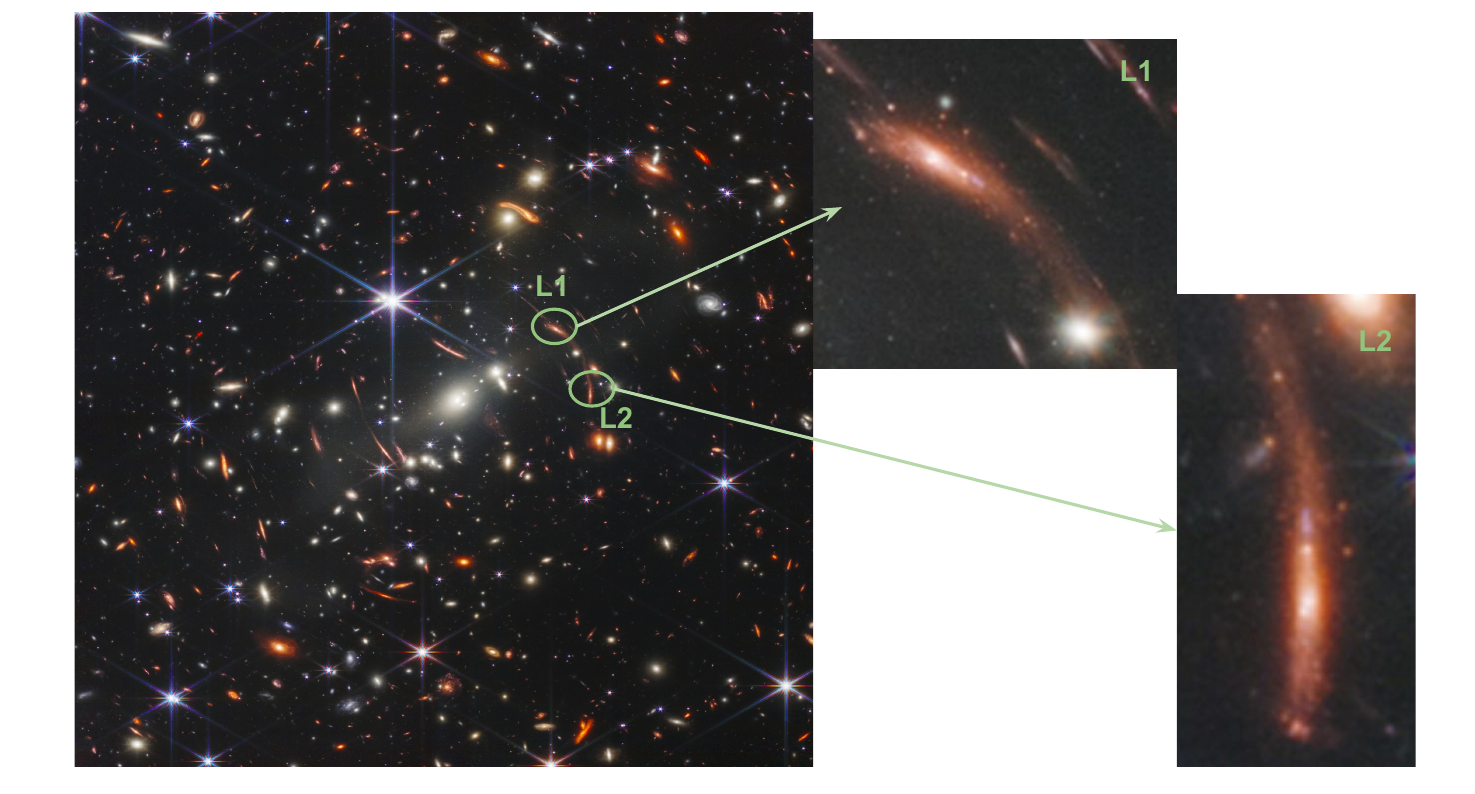}
    \caption{The lens cluster \textsc{SMACS0723} field imaged by JWST. Here we highlight the lens arc L1 and L2 respectively, where the point-alike sources around the main lens arc provides extra information, thus offering the opportunity to apply our algorithm.}
    \label{fig: jwst-lens-field}
\end{figure*}

Despite the huge success of the $\Lambda$-cold dark matter ($\Lambda$CDM) model in explaining the universe on large scales, an elementary understanding of dark matter (DM) is still lacking. 
The internal structures of DM haloes provide important insights into the properties of DM, particularly regarding their interactions on scales smaller than the typical halo size. 
Self-interactions among the dark matter `particles', as suggested by some DM theories, could play a role in addition to self-gravitation, modifying the equilibrium configuration of the haloes. 
Phenomenological problems, such as the cusp-core problem \citep{cuspy-core-dwarf-galaxy-nature1994,obs-contraints-on-singular-halo,cusp-core-review} and, to some extent, the missing satellite problem \citep{missing-satellite-original1999,falisification-of-cdm-from-galaxy,plane-of-satellites-problem,small-scale-challenge-to-cdm-annurev,missing-satellites-review2010}, can be explained by this type of model \citep{wavedm-sim-tom} without invoking more complicated modelling of baryonic processes such as those implemented in \citep{cusp-core-baryon-sol} (see also \citet{missing-satellite-cdm-solution} for an extensive review).

In addition to the configuration of the main halo distribution, more localised structures within the halo have also been identified. 
The existence of DM substructures, such as sub-haloes that orbit around the main halo, has been confirmed by CDM simulations \citep{cdm-subhalo-population-sim,subhalo-mf-semi-analytic}. 
Beyond the CDM model, alternative DM models such as fuzzy/wave dark matter \citep{wave-dm-review-lam-hui,wavedm-sim-tom,subhalo-problem-fuzzyDM} predict haloes in dynamical equilibrium, with approximately Gaussian/Gamma random density fluctuations oscillating within the halo \citep{oguri-waveDM-subhalo-gaussian-random-field-approx}.

The identification of DM haloes based on strong gravitational lensing has been studied and applied for around a decade. 
The strong lensing of quasars, and the inconsistency in their signed sum of magnifications across multiple images (usually known as flux ratio anomalies) has been proposed as a probe of DM sub-haloes \citep{madau-quasar-lens-substructures}. 
Such observational campaigns have been performed regularly to sample as many quasar lens systems as possible, since statistical stacking of such anomalies is necessary to draw robust conclusion about the properties of the underlying sub-halo population \citep{flux-anomalies-sub-structure-statistics-analysis,cdm-substructure-flux-anomal-gilman17,self-interacting-dm-flux-anomal-gilman21}.
In the second half of the 2000s, there was another major advance in the search for individual sub-halo es rather than the statistical population, using the technique `gravitational imaging' applied to galaxy-scale strong lenses \citep{Koopmans-2005-gravitational-imaging,lens-substructure-vegetti2008}. 
This was achieved by sophisticated forward modelling, and advanced computational techniques such as tessellation schemes \citep{vegetti-sub-halo-detection-nature,suyu-source-inversion}, regularised optimisation methods (as reviewed in \citealt{dm-substructure-strong-lensing-vegetti-2023-review}), chained inference methods \citep{lensing-chained-inference-etherington2022} and pseudo-linear inversion \citep{semi-linear-lens-inversion-2003,semi-linear-inversion-dye2005,nightingale-pseudo-linear-inversion} , developed over the last decade to improve convergence in sophisticated forward modelling. 
Indeed, there have been remarkable advances in `gravitational imaging' \citep{durham-forward-modelling-lensing-dm-mass,hezaveh-alma-substructure-detection}. 
Using the CDM sub-halo model as a prior, detections of sub-haloes as light as $M \sim 10^8\,M_\odot$ have been reported \citep{vegetti-sub-halo-detection-nature}. 
It is also worth mentioning the recent claimed discovery of a $M \sim 10^6\,M_\odot$ sub-halo using very long baseline interferometry \citep{vegetti-vlbi-lightsubhalo-profile}, and its physical interpretation was explored in an accompanying paper \citep{vegetti-lightsubhalo-interpretation-in-TNGsim}.

Among most of the claimed sub-halo detections so far, it has been found that the best fit sub-halo model deviates from the CDM prediction \citep{overconcentrated-halo-minor2021,angular-model-alma-overconcentration,aris-multipole-perturbation-subhalo-overconcentration}, featuring an over-concentrated core substantially different from the expectation of the standard mass–concentration relation (MCR) \citep{mass-concentration-relation-2007}. 
Direct profiling of the sub-haloes from observational data appears difficult, as the analysis methods are often based on parametric lens models, followed by a regularised and pixelated scheme to capture the perturbations on top of the parametric lens model. 
The introduction of extra regularisation hyper-parameters is inevitable in these methods to avoid under-constrained model fitting. 
In particular, recent work \citep{qiuhan-lenslight-mcr-solution} has suggested that the over-concentration problem can be partially resolved by incorporating more sophisticated modelling of the lens light, so that the excess flux detected from a narrow region in the lens arc can be interpreted as arising from the `under-subtracted' light from the lens plane, rather than the steepening of the gravitational potential. 
This highlights a key limitation in the existing methods based on parametric forward modelling: detection of sub-haloes is strongly dependent on the specific parametric model, and the use of a more detailed model might be able to explain away the previously found sub-halo signatures.

Advanced modelling techniques to capture the realistic main lenses (which host the sub-halo lenses) exist and attempt to include more variation in the macroscopic lens model \citep{aris-multipole-perturbation-subhalo-overconcentration,sam-multipole-vs-substructure}, or even to use the so-called `free-form' modelling \citep{xiaoyeue-freeform-lens-modeling}. 
They aim to capture a more generic and realistic model, so as to examine if the sub-halo signal could be absorbed by a more complicated macroscopic lens model. 
Attempts to quantify other parameter degeneracies \citep{subhalo-detection-data-quality-assessment,enzi-subhalo-systematics-sim} are usually based on simulated observations, and the assumed complexity in the injected model sets the bottleneck and limits the scope of this kind of study. 

It is clear that one direction is to incorporate more complicated lens models by introducing more parameters, which further complicates efforts to characterise the degeneracies in the already large parameter space. 
There are still unexplored gaps in understanding the parameter degeneracies involved in such a huge parameter space, typical of forward modelling, let alone assessing the robustness of the claimed sub-halo detection.

As an attempt to bypass the aforementioned challenges in parametric forward modelling, we show in this paper that the presence of DM substructures — broadly including CDM sub-haloes and other localised small-scale departures from the smooth halo model — can be robustly identified without any presumed parametrisation, neither lens model nor the source plane model. 
This is based on an elementary law that is valid in geometrical optics: the \textit{optical Liouville theorem}. 
The specific corollary of the optical Liouville theorem — that the flux magnification is equivalent to the size magnification, which we refer to as the \textbf{geometrical duality} — is the central idea behind our proposed method.

Anomalies in observed flux — as in the classical observations of multi-imaged lensed quasars — are one of the typical ways to argue for substructures. 
The non-parametric anomalies metric — the normalised signed-sum of flux magnifications — is predicted to be zero for various universal asymptotic forms (fold, cusp, and the higher order caustics) of lens models with the aid of catastrophe theory \citep{catastrophe-arnold-ade-classification,gravitational-lensing-book-with-singularity-schneider1992,singularity-lensing-petters-book-2001}. 
The validity of such a metric relies on the global information of the underlying gravitational potential, and localised perturbations, such as additional shear by nearby galaxies \citep{external-shear-anomaly-resolve}, also modify the metric. 
In general, there is no way to distinguish whether the flux anomalies are generated by DM substructures or other types of gravitational perturbations (such as the fuzzy DM granules \citet{waveDM-caustic-crossing,wavedm-lensing-amruth})  that change only one of the multiple images via the scalar anomaly parameter.

The use of geometrical duality as proposed in this paper, however, fully utilises the local information available, therefore enabling the separation between the perturbations generated by substructures and other types of perturbation that operate on the mesoscale. 
The working principle of the algorithm is generic, but the practical aspect of applying this approach to analyse gravitational lenses was not made possible until the commissioning of the James Webb Space Telescope (JWST) \citep{jwst-first-light}. 
This is due to the improvement in limiting magnitude available on JWST, as well as its unique infrared imaging capability. 
These improvements allow for resolving point-alike sources around the spatially extended lens arcs. 
The point-like sources form a `constellation', enabling the construction of non-parametric lens observables to quantify the local lens geometry. 
More generically, some other non-parametric observables have been studied, and they are comprehensively reviewed in \citep{review-model-independent-strong-lensing}, with their physical implications thoroughly discussed. 
These other observables are mostly based on asymptotic behaviours and symmetry of the underlying lens potential, which are vastly different from what we propose here, which is purely from geometrical optics.

In Fig.~\ref{fig: jwst-lens-field}, we show the lens field SMACS0723 imaged by JWST, on which the proposed algorithm is applied. 
Throughout the paper, we assume flat-$\Lambda$CDM cosmology with $H_0 = 70$ km s$^{-1}$ Mpc$^{-1}$ and $\Omega_m = 0.3$ solely for the purpose of obtaining different cosmological distances. 
Despite the CDM cosmology being assumed for cosmological distances, we emphasise that we do not need to assume any specific DM model in the analysis of localised haloes. 
As a result, the following discussions are generic unless specified otherwise, mainly in the case of comparison with the existing literature.

\section{Gravitational Lensing and Conservation Theorems}
To arrive at a model-independent algorithm to detect dark matter sub-structures, it is important to utilise the inherent conservation law respected by the general lens equation - rather than studying the solutions of the lens equation under specific gravitational potentials. 
For this purpose, we review the basics of gravitational lensing below — with an emphasis on the conservation law and their observational consequences for astronomical observations.

\subsection{Gravitational Lensing}
The General Theory of Relativity predicts the deflection of light rays under the influence of gravitational potential. 
The gravitational deflection angle, in the Newtonian limit, reads:
\begin{equation}
 \alpha_i(\theta) = g(z_\ell, z_s) \partial_i \Phi(\theta),
\end{equation}
where $g(z_\ell, z_s)$ is the lensing efficiency that depends on the angular diameter distances of the source at $z=z_s$ and the lens at $z=z_\ell$. 
The 2D projected gravitational potential $\Phi(\theta)$ in the above expression is:
\begin{equation}
 \Phi(\theta) \equiv \int_{z_\ell}^{z_\ell + \delta_z} \frac{\partial r}{\partial z'}  \,\phi(\theta,z') dz',
\end{equation}
which is an integral that projects along the line-of-sight thickness of the gravitational lens. 
These equations are the only parts of lensing theory that are specific to gravitation. 
Practically, in most observational settings, the lensing efficiency $g(z_\ell, z_s)$ can be determined independently using spectroscopy or photometry, therefore reducing the efficiency factor to an overall scaling constant for the lens potential. 
For this reason, we drop the lensing efficiency in the following discussions. 
In all the numerical results shown in this work, the lensing efficiency has been correctly accounted for.

\subsection{Classical Geometrical Optics} \label{sect: classic-optics-review}
With the deflection angle $\alpha(\theta)$ defined by the theory of gravitation, the remaining description of lensing effects is purely geometrical. 
It is convenient to think about the effect of lensing - instead of the propagation of light rays - as a transformation law that relates the \textbf{image plane} and the \textbf{source plane}. 
We denote the image-plane coordinates with 2D angular vectors $\theta_i$ and, likewise, the source-plane coordinates $\beta_i$. 
The transformation between the two coordinate systems is specified by the propagation of light rays, often called the lens equation:
\begin{equation} \label{eq: lens-eq}
 \beta_i = \theta_i - \alpha_i(\theta).
\end{equation}
Consider the transformation of an infinitesimal element $\delta \theta$ that we observe on the image plane to the underlying (and unobserved) source element with size $\delta \beta$. 
The size transformation is given by:
\begin{equation}
\begin{split}
 \delta \beta_i &= \frac{\partial \beta_i}{\partial \theta_j} \delta \theta_j \\
 &= \left( 1_{ij} - g(z_\ell, z_s)\frac{\partial^2 \Phi}{\partial \theta_i \partial \theta_j}\right) \delta \theta_j.
\end{split}
\end{equation}
On the second line we used the lens equation (Eq.~\ref{eq: lens-eq}). 
Therefore, an observed image with angular size $\delta \theta_x \delta \theta_y$ corresponds to the actual source of size:
\begin{equation}
\mu_{\rm size} = \frac{\
\delta \theta_x \delta \theta_y\
}{\
\delta \beta_x \delta \beta_y\
} = \mathrm{det}\left( 1_{ij} - g(z_\ell, z_s)\frac{\partial^2 \Phi}{\partial \theta_i \partial \theta_j} \right)^{-1},
\end{equation}
and this holds in the infinitesimal sense. 
This defines the size magnification of a lens system.

\subsection{Transformation of surface brightness}
In practice, the size of an object is implicitly defined via its surface brightness distribution $\mathcal{I}(\theta)$. 
Consequently, we should also consider the transformation of $\mathcal{I}(\theta)$ under the action of lensing, rather than examining only the coordinate change ($\theta \rightarrow \beta$). 
The observed surface brightness (in the image plane) is determined by both the lens configuration and the intrinsic source brightness distribution $\mathcal{I}_{\rm src}(\beta)$. 
Suppose that the image transformation from $\theta$ to $\beta$ is an isomorphism: i.e. it is invertible and differentiable in (at least) some compact subspace.
\footnote{In strong lensing, this condition is true with the exception along the critical curve (equivalently the caustics on the source plane).}
With finite resolution limited by diffraction, the observed surface brightness is obtained by integrating over a discretised interval with width equivalent to the pixel scale. 
Suppose that a pixel on the image plane that spans the interval $\Theta = [\theta, \theta+\Delta \theta]$ is mapped to the source plane $[\beta, \beta+\Delta \beta]$. 
The observed, pixelated surface brightness at pixel $\Theta$ is:
\begin{equation} \label{eq: brightness-transform}
\begin{split}
 \mathcal{I}(\Theta) &= \frac{1}{\Delta \theta}\int^{\theta+\Delta \theta}_{\theta} \mathcal{I}_{\rm src}(\beta(\theta')) \,d\theta' \\
 &= \frac{1}{\Delta \theta}\int^{\beta+\Delta \beta}_{\beta} \mathcal{I}_{\rm src}(\beta') \frac{\partial \theta}{\partial \beta'} \,d\beta',
\end{split}
\end{equation}
where on the second line we express the integration domain in terms of the source coordinate, and thus include the extra Jacobian factor $\partial \theta/\partial \beta'$ for the covariance of $d\beta'$. 
If the pixel size $\Delta \theta$ is small enough compared to $\Delta(\partial \beta/\partial \theta)$, the integration limits on the second line can be rewritten as a contravariant transformation: 
$\Delta \beta = (\partial \beta/\partial \theta)\,\Delta \theta$. 
As such, we can now consider one of the two limiting cases of source light distribution. 
If $\mathcal{I}_{\rm src}(\beta)$ remains almost constant $\mathcal{I}_0$ within a pixel, the integral reads:
\begin{equation} \label{eq: conserve-surface-brightness}
 \mathcal{I}(\Theta) \approx \frac{\partial \theta}{\partial \beta}\frac{\partial \beta}{\partial \theta} \mathcal{I}_0 = \mathcal{I}_0.
\end{equation}
One can appreciate that the surface brightness remains constant under gravitational lensing. 
This is a manifestation of the optical Liouville theorem \citep{optical-liouville-theorem-gr-proof-thorne}. 
However, because the image size of an object with intrinsic size $\Delta_\beta > \Delta \beta$ is magnified into an image with size $\Delta_\theta \equiv \mu_{\rm size}\Delta_\beta$, the total flux received by the observer is enhanced as $\mathcal{I}_0 \Delta_\theta = \mu_{\rm size}\mathcal{I}_0 \Delta_\beta$.

\subsection{Optical Liouville Theorem}
While the mathematical proof of the optical Liouville theorem can be found in standard textbooks, and particularly in the context of general relativistic spacetime in \citep{optical-liouville-theorem-gr-proof-thorne}, for ease of discussion, we focus on the intuitive aspect of it and, most importantly, its consequence for our application. 
The optical Liouville theorem can be conceptualised by considering a bundle of light rays. 
Upon interaction with any passive optical element (for example, gravitational lenses), the direction of ray propagation will be altered. 
This alteration - including the possibility of redirecting and focusing more rays originating from a larger aperture - gives rise to more rays arriving at the destination (i.e. an increment in flux). 
However, the size of the cross-section of such an altered bundle of rays also increases accordingly (i.e. an increment in size). 
The optical Liouville theorem states that the focusing power and the enlargement of the bundle cross-section are dual to each other. 
As such, an obvious corollary is:
\footnote{Another corollary of the theorem is the conservation of surface brightness, which has been discussed implicitly in Eq.~\ref{eq: conserve-surface-brightness}.}
\begin{equation} \label{eq: geo-duality}
 \mu_{\rm size} = \mu_{\rm flux},
\end{equation}
in the infinitesimal sense. 
This equivalence is the central object studied in this paper, and we refer to this relation as \textbf{geometrical duality}. 
The observational aspect of geometrical duality is intriguing: in practice we measure these quantities by summation over intervals of finite size (i.e. pixels). 
Most importantly, the size magnification can be measured only up to the diffraction limit, while the flux magnification can be measured up to the photon-counting precision, and thus well beyond what the diffraction limit allows. 
This is the foundation for achieving super-resolution - that is, retrieving information beyond the diffraction limit from gravitational lenses. 
In particular, we are interested in the specific case where the gravitational lenses consist of dark matter sub-structures with characteristic scales below the diffraction limit of existing optical telescopes. 
Finally, geometrical duality is purely an optical theorem, and thus independent of the specific configuration of the gravitational lens. 
In the following, we exploit this fact to outline a model-independent algorithm to detect dark matter sub-structures.

\section{Algorithm} \label{sect: algo}
\subsection{Breakdown of Geometrical Duality and DM Sub-structures}

We have discussed the action of lensing on pixelated surface brightness - in the limit of a smooth source with constant brightness across (Eq.~\ref{eq: conserve-surface-brightness}). 
Those results resemble the standard conclusion of geometrical duality (Eq.~\ref{eq: geo-duality}). 
At the opposite extreme, where the source is an infinitesimal point with $\mathcal{I}_{\rm src}(\beta) = \mathcal{I}_0 \delta_D(\beta-\beta_0)$ being a Dirac-delta, 
intriguing effects follow. 
The transformation of the integration interval $\Delta \theta \rightarrow \Delta \beta$ in Eq.~\ref{eq: brightness-transform} does not affect the integration result of the Dirac-delta as long as the interval covers $\beta_0$. 
In such case:
\begin{equation}
 \mathcal{I}(\Theta) = \frac{\partial \theta}{\partial \beta}\bigg|_{\beta=\beta_0}\mathcal{I}_0 = \mu_{\rm size}\mathcal{I}_0 ,
\end{equation}
In practice, any realistic source can be written as a combination of a constant/smooth source profile with the addition of smaller features represented by a summation of Dirac-deltas. 
And therefore, the pixelated surface brightness $\mathcal{I}(\Theta)$ would also encapsulate the information of $\mu_{\rm size}$, provided that the source possesses brightness structures smaller than the pixel scale $\mu_{\rm size} \Delta_\beta < \Delta \theta$. 
We wish to highlight here that geometrical duality (Eq.~\ref{eq: geo-duality}) is apparently violated in this case with pixelated observation. 
As the size-magnified object is not resolved, we would incorrectly infer that there is no size magnification ($\mu_{\rm size} = 1$), yet the flux magnification associated with this point can easily be detected ($\mu_{\rm flux} \neq 1$). 
The observed violation of geometrical duality ($\mu_{\rm size} \neq \mu_{\rm flux}$) thus suggests two possibilities, namely:
\begin{enumerate}
 \item The source brightness distribution possesses small-scale features; and/or
 \item The lens potential - and consequently the lensing magnification - varies on scales smaller than the resolution limit. This is the case in the presence of dark sub-structures.
\end{enumerate}

If we exclusively focus on analysing the point sources as identified in observations, we can rule out the first case and consequently \textbf{attribute all observed violations of geometrical duality to the presence of dark sub-structures}. 
To see this, consider the Einstein radius of a point-mass-like halo, which scales as (assuming $z_s = 2.2$ and $z_l = 0.39$, which is the configuration for the lens system shown in Fig.~\ref{fig: jwst-lens-field}):
\begin{equation} \label{eq: einstein-radius}
\begin{split}
 \theta_E &= \left( \frac{4GM}{c^2} \frac{D_{ls}}{D_s D_l} \right)^{1/2} \\
 &= 0.0232\left(\frac{M}{10^8 M_\odot}\right)^{1/2}\mathrm{arcsec},
\end{split}
\end{equation}
In the presence of a macroscopic gravitational gradient implemented by the host DM halo, \citep{oguri-caustic-crossing-analytic} showed that the effective Einstein radius of the sub-halo is enlarged by a factor given by either the radial or tangential eigenvalue of $\partial \beta_i/\partial \theta_j$ (as defined in Eq.~\ref{eq: lens-eq}), depending on the image parity. 
For the purpose of quick estimation, we approximate this factor as $\sqrt{\mu_{\rm size}}$, so the enlarged Einstein radius of the sub-halo is of order $\sqrt{\mu_{\rm size}}\,\theta_E$. 
Considering a nominal case where a $M_{\rm sub} \sim 10^7 M_\odot$ sub-halo is placed under a background magnification of $\mu_{\rm size} \approx 100$, the corresponding Einstein radius would be of order $\mathcal{O}(10^{-3})$ arcsec, which is substantially smaller than the diffraction limit of JWST. 
As such, the spatial effect of this sub-halo cannot be observed directly from the imaging data. 
However, the magnification effect can still be seen in terms of flux magnification $\mu_{\rm flux}$ thanks to the optical Liouville theorem. 
This remains valid as long as the source light that illuminates the sub-halo lens closely resembles a point source.

As long as we can derive the macroscopic magnification factor (which is $\mu_{\rm size} = 100$ in this example), we can \textbf{isolate} the magnification produced by the $M_{\rm sub} \sim 10^7 M_\odot$ sub-halo, thus delivering constraints on this class of unseen DM sub-structures. 
In the following, we describe how this can be done.

\subsection{Geometrical Duality in a Collection of Point Sources}

If we restrict ourselves to analysing point-alike sources in lensing systems, the violation of geometrical duality — namely that the excessive fluxes observed from the sources indicate the existence of excessive gravitational structures such as DM sub-haloes. 
These excessive gravitational structures are compositely contributed by macroscopic component that is larger than the pixel scale, and a microscopic component that is smaller than the pixel scale. 
For the detection of sub-structures, we separate the microscopic contribution from the macroscopic component. 
This can be done by considering a collection of point-alike sources in which the separations amongst the point sources are resolved. 
We refer to such collections as \textbf{constellations}.

For the sake of explanation and without loss of generality, we shall focus on the example of a dipole of one-dimensional point sources in this section, while realistic two-dimensional lensing will be discussed later when we apply our algorithm to the JWST data. 
Suppose the one-dimensional gravitational magnification $\mu_{\rm size}(\theta)$ consists of a macroscopic component and a small, localised structure, so that:
\footnote{Note that this reciprocal sum of magnification can only be justified in 1D lensing; in 2D, similar decomposition must be generalised to a sum of the deformation matrix $\partial \beta/\partial \theta$, so that the Jacobian determinant of the resulting deformation matrix would inform the magnification.}
\begin{equation}
 \mu_{\rm size}^{-1}(\theta) = \mu_0^{-1} + \mu_{\rm sub}^{-1} \delta_D(\theta-\theta_0),
\end{equation}
and we are probing such a configuration using a dipole of point sources, with the surface brightness distribution in the source plane given by:
\begin{equation}
 \mathcal{I}(\beta) = I_1 \delta_D(\beta-\beta_1) + I_2 \delta_D(\beta-\beta_2),
\end{equation}
so that the two point sources are separated by a distance of $\,\beta_1 - \beta_2\,.$

If we choose to have a perfect overlap of the Dirac-delta lens with the first point source, $\theta_0 = \mu_{\rm sub}\beta_1$. 
The resulting configuration under the action of the gravitational magnification is:
\footnote{In realistic 2D lensing, such a simplified formula does not hold: the total magnification on $I_1$ is a nonlinear combination of the eigenvalues that contribute to $\mu_0$ and $\mu_{\rm sub}$. However, the idea that the separation between $I_1$ and $I_2$ traces only $\mu_0$ while the flux traces both $\mu_0$ and $\mu_{\rm sub}$ remains valid.}
\begin{equation}
 \mathcal{I}_{\rm lensed}(\theta) = 
 (\mu_0^{-1} + \mu_{\rm sub}^{-1})^{-1} I_1 \delta_D(\theta - \mu_0\beta_1)
 + \mu_0 I_2 \delta_D(\theta - \mu_0\beta_2).
\end{equation}

It is clear that the separation between the two sources is now $\mu_0\,|\beta_1 - \beta_2|$, which is the expected macroscopic magnification in size. 
Intriguingly, the microscopic component of the gravitational structure $\mu_{\rm sub}$ does not contribute to any apparent change in size, but manifests only as a flux magnification of $I_1$.

To extract such gravitational structures that vary on scales smaller than the pixel scale, one can therefore estimate the size magnification by the separation of the dipole, and subsequently remove this macroscopic contribution from the measured flux of the point sources. 
The residual excessive flux \textbf{can only be attributed to} such a gravitational sub-structure.

\subsection{Verification of Geometrical Duality}

In practice, the source plane is never observed and is thus subject to extra modelling uncertainty. 
However, it is possible to bypass the unnecessary source-plane modelling and make the method fully data-driven. 
The trick is to consider strongly lensed systems in which point-source constellations are multiply imaged along different lens arcs. 
Such systems are now readily identified, with their physical origin remaining uncertain. 
The most compelling explanation to date invoke a galaxy with massive proto-star clusters located on its outskirts.

With pairs of multiply imaged constellations, it is possible to infer the image-to-image magnification instead of the image-to-source magnification, which allows for a fully data-driven approach without model dependence. 
As the transformation among different regions on the image planes can be understood as a composition of a first-image-to-source transformation and a source-to-second-image backward transformation, the optical Liouville theorem and its corollaries also apply to the image-to-image transformation. 
The unobserved intrinsic brightness of the point sources can be factored out by constructing flux ratios from multiple images, so that only gravitational effects remain to be estimated.

The overall approach for detecting gravitational sub-structures in a model-independent way would thus be:
\begin{enumerate}
 \item Identify a pair of multiply lensed systems of point-source constellations, most likely discoverable in cluster lenses.
 \item Pair the point sources on the multiple images, using the light distribution of the host galaxy as a visual guide to assist the pairing.
 \item Measure the flux magnification ratios of each of the point-source pairs.
 \item Measure the size magnification ratio between the separations among the point sources.
 \item If the measured size magnification does not agree with the flux magnification, there is likely gravitational sub-structure lying on that specific point source.
\end{enumerate}

Step (iv), however, requires extra care, as realistic lensing effects are observed on the two-dimensional celestial sphere. 
In the following section, we construct the size magnification estimators in 2D and assess the robustness of such estimators with numerical simulations.

\section{Direct Estimation of Size Magnification in 2D} \label{sect: estimators}

\subsection{Construction of Estimators}

\begin{figure*}
 \centering
 \includegraphics[width=\textwidth]{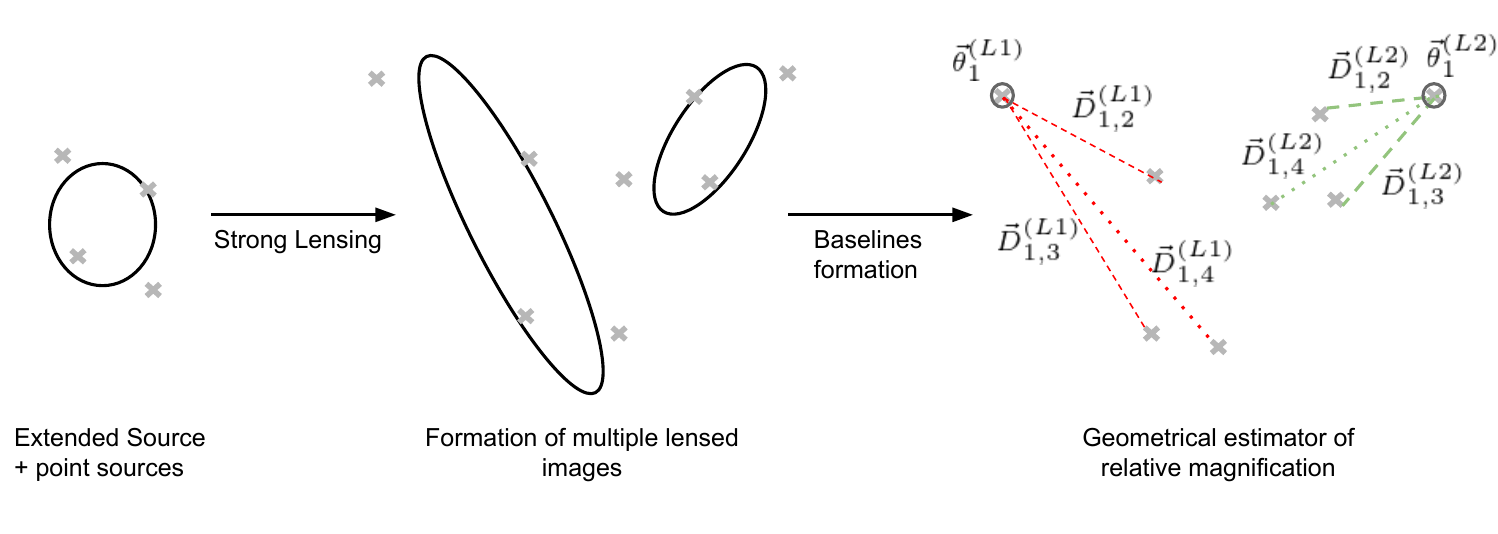}
 \caption{Overview of the algorithm. The mathematical notation for geometrical quantities are defined as above on the rightmost panel.}
 \label{fig: algo-notation-overview}
\end{figure*}

\subsubsection{Framework and Notations}

We aim to design an algorithm that explicitly checks for the validity of geometrical duality without assuming any model. 
This requires constructing estimators directly from observables. 
We therefore study the relative transformation amongst the observed multiple images, instead of back-tracing to the source plane as in standard forward-modelling approaches.

In the case of a pair of neighbouring lens arcs (which is typical in the fold configuration), the relative magnification amongst the multiple images is typically of order $\sim\mathcal{O}(1)$, 
which is advantageous for minimising the influence of extreme deformation.

Consider the image pairs $\vec{\theta}^{(L1)}_a$ and $\vec{\theta}^{(L2)}_a$ that correspond to the same source coordinate $\beta_a$. 
Here, the subscript $a$ runs over the multiply imaged point-source pairs. 
The two coordinate systems can be related using the lens equation (Eq.~\ref{eq: lens-eq}):

\begin{equation}
\begin{split} \label{eq: two-image-lens-eq}
 \vec{\beta}_a = \vec{\theta}^{(L1)}_a - \nabla\Phi\big|_{(L1),a} 
               &= \vec{\theta}^{(L2)}_a - \nabla\Phi\big|_{(L2),a} , \\
 \vec{\theta}^{(L1)}_a - \vec{\theta}^{(L2)}_a 
    &= \nabla\Phi\big|_{(L1),a} - \nabla\Phi\big|_{(L2),a} ,
\end{split}
\end{equation}

which states that the separation of the two images is directly related to the difference in the lens-potential gradients, evaluated at the two images respectively. 
This may also be interpreted as the transformation law that relates the two images.

Denote this transformation from L2 to L1 as $\zeta$ which is a function of $\vec{\theta}^{(L2)}$, the transformation law can be written as:
\begin{equation}
 \vec{\theta}^{(L1)}_a = \vec{\theta}^{(L2)}_a + \zeta\!\left(\vec{\theta}^{(L2)}_a\right) .
\end{equation}

It is possible to Taylor expand $\zeta$ as long as the transformation remains one-to-one and infinitely differentiable.
This allows us to relate between one specific point source $_a$ to the another $_b$. 
\footnote{This condition holds because we explicitly ensure that the image plane is segmented into regions that do not cross critical curves. 
Our simulations confirm that this expansion yields accurate results.}
The leading-order terms are:
\begin{equation}
 \zeta\!\left(\vec{\theta}^{(L2)}_b\right) 
   = \zeta\!\left(\vec{\theta}^{(L2)}_a\right)
   + \frac{\partial\zeta}{\partial\vec{\theta}^{(L2)}}\bigg|_a
     \left(\theta^{(L2)}_b - \theta^{(L2)}_a\right)
   + \mathrm{h.o.},
\end{equation}
where h.o. denotes higher-order terms.

We now consider the finite differences within the same lens arc, called the \textbf{baselines}:
\begin{equation}
 \vec{D}^{(L2)}_{a,b} = 
   \vec{\theta}^{(L2)}_{b} - \vec{\theta}^{(L2)}_{a}.
\end{equation}

Using this definition, the transformation from L2 to L1 becomes:
\begin{equation} \label{eq: baselines-relation}
 \vec{D}^{(L1)}_{a,b}
   = \vec{D}^{(L2)}_{a,b}
   + \frac{\partial\zeta}{\partial\vec{\theta}^{(L2)}}\bigg|_a
     \vec{D}^{(L2)}_{a,b}
   + \mathrm{h.o.}
\end{equation}

Both $\vec{D}^{(L2)}_{a,b}$ and $\vec{D}^{(L1)}_{a,b}$ are directly measurable from the data (see Fig.~\ref{fig: algo-notation-overview}). 
With $\ge 2$ baselines per point, we can measure the transformation tensor 
$\partial\zeta/\partial\vec{\theta}^{(L2)}$, represented by a $2\times2$ matrix.

This tensor is also an estimator of the relative geometrical magnification:
\begin{equation}
 \mu_{\rm size}(a)
   \equiv 
   \mathrm{det}\!\left(
      \frac{\partial\vec{\theta}^{(L1)}}{\partial\vec{\theta}^{(L2)}}\bigg|_a
   \right)
   = 
   \mathrm{det}\!\left(
      \mathbf{1}
      + 
      \frac{\partial\zeta}{\partial\vec{\theta}^{(L2)}}\bigg|_a
   \right),
\end{equation}
where $\mathbf{1}$ is the $2\times2$ identity matrix.

Thus, we obtain a general framework for computing the relative geometrical magnification directly from the data, without assuming any lens model.

\subsubsection{2v Jacobian}

Generally, estimating the $2\times2$ matrix 
$\partial\zeta/\partial\vec{\theta}^{(L2)}$ 
requires four independent constraints, 
corresponding to measuring two 2D baselines 
$\vec{D}_{a,b}$ and $\vec{D}_{a,c}$ 
formed by three point sources $\{a,b,c\}$ along the same lens arc.

The estimation error increases systematically as $\|\vec{D}\|^2$, 
therefore the seemingly obvious choice is to use points $b$ and $c$ that are closest to point $a$, 
thus minimising 
$\|\vec{D}^{(L2)}_{a,b}\|$ 
and 
$\|\vec{D}^{(L2)}_{a,c}\|$.

After selecting $b$ and $c$, the transformation tensor can be estimated via linear least squares. 
Define:
\begin{equation} \label{eq: linear-estimator-definition}
 X_a \equiv 
 \begin{pmatrix}
   \vec{D}^{(L2)}_{a,b} \\
   \vec{D}^{(L2)}_{a,c}
 \end{pmatrix},
 \qquad
 Y_a \equiv
 \begin{pmatrix}
   \vec{D}^{(L1)}_{a,b} - \vec{D}^{(L2)}_{a,b} \\
   \vec{D}^{(L1)}_{a,c} - \vec{D}^{(L2)}_{a,c}
 \end{pmatrix}.
\end{equation}

The estimator is then:
\begin{equation}
 \frac{\partial\zeta}{\partial\vec{\theta}^{(L2)}}\bigg|_a 
   \approx 
   X_a^{-1} Y_a.
\end{equation}

Although this expression appears algebraically heavy, 
geometrically it corresponds to comparing the area of the triangle formed by $\{a,b,c\}$ in L2 to that in L1. 
We therefore refer to this estimator as the \textit{2v Jacobian}.

\subsubsection{Quad Jacobian}

While the \textit{2v Jacobian} works reasonably well, 
it has inherent weaknesses. 
In particular, the shortest baselines in L2 may not correspond to the shortest baselines in L1, 
because the transformation must be invertible
\footnote{The transformation must have non-zero Jacobian determinant as long as the domain does not cross a critical curve.}. 
This creates inconsistencies in the estimation procedure.

Because the choice of baselines is arbitrary, 
a more robust approach is to average information from \textbf{all} available baselines. 
Examining Eq.~\ref{eq: baselines-relation}, baselines with larger 
$\|\vec{D}^{(L2)}\|$ 
contribute more to higher-order errors, scaling as 
$\|\vec{D}^{(L2)}\|^2$.

To down-weight long baselines, define the weighting matrix:
\begin{equation}
 W_a \equiv
 \begin{pmatrix}
   1/\|\vec{D}^{(L2)}_{a0}\|^2 \\
   \vdots \\
   1/\|\vec{D}^{(L2)}_{aN}\|^2
 \end{pmatrix}.
\end{equation}

We then solve the weighted least-squares system:
\begin{equation}
 (X_a^T W_a X_a)\,
 \frac{\partial\zeta}{\partial\vec{\theta}^{(L2)}}\bigg|_a
   = 
 X_a^T W_a Y_a ,
\end{equation}
where $X_a$ and $Y_a$ extend Eq.~\ref{eq: linear-estimator-definition} to include all non-trivial baselines. 
The optimal solution is obtained via singular value decomposition (SVD). 

Because this estimator uses multiple baselines with quadratic weighting, 
we refer to it as the \textit{Quad Jacobian} (or QW Jacobian, which stands for quadratically-weighted).

\subsection{Systematic Errors of the Estimators}

In order to validate the magnification estimators derived above, we performed a simulation to assess the accuracy of the estimators in recovering the injected values. 
The detailed setup and analysis of the simulation are presented in Appendix~B. 
Here, we highlight the key takeaways from the results.

First of all, as a generic behaviour, both the 2v Jacobian estimator and the Quad Jacobian estimator can successfully recover the underlying magnification ratios across multiple images. 
The performance improves with the number of point sources available on the source plane, so that the typical separation between neighbouring point sources on the lens arc is as compact as possible. 
Theoretically, this is evident from the fact that our estimators are constructed via a Taylor expansion, which relies on the local linearity of the lensing action. 
The typical fractional error in recovering the underlying magnification ratio is $\sim \pm 10\%$.

However, the performance does not show scaling with the relative magnification across the lens arcs. 
This is favourable, as the absence of such scaling implies that there is no significant bias to be corrected when a population of various lens-arc pairs is to be analysed in the future.

Across the two kinds of estimators — the 2v Jacobian and the Quad Jacobian — the Quad Jacobian recover the underlying magnification ratios without predicting extreme outliers, in contrast to the 2v Jacobian. 
In fact, there is a clear trade-off: while the 2v Jacobian uses the more localised information without considering the point-source pairs further apart, it is more unstable, as the fractional uncertainty in inferring the baseline distance $\vec{D}_{a,b}$ is larger because of the limited ability to localise the point sources. 
The Quad Jacobian is better in this regard, as non-local information is used to perform weighted averaging of the baselines.

Yet, this procedure of including long baselines also introduces a bias because it fails to capture higher-order spatial variation in the lensing magnification. 
Consequently, the Quad Jacobian shows a tendency to use the average magnification across the entire lens arc and to under-estimate the local differential magnification, albeit a small effect.

\section{Proof of Concept in SMACS0723} \label{sect: poc}
\begin{figure*}
    \centering
    \includegraphics[width=.9\textwidth]{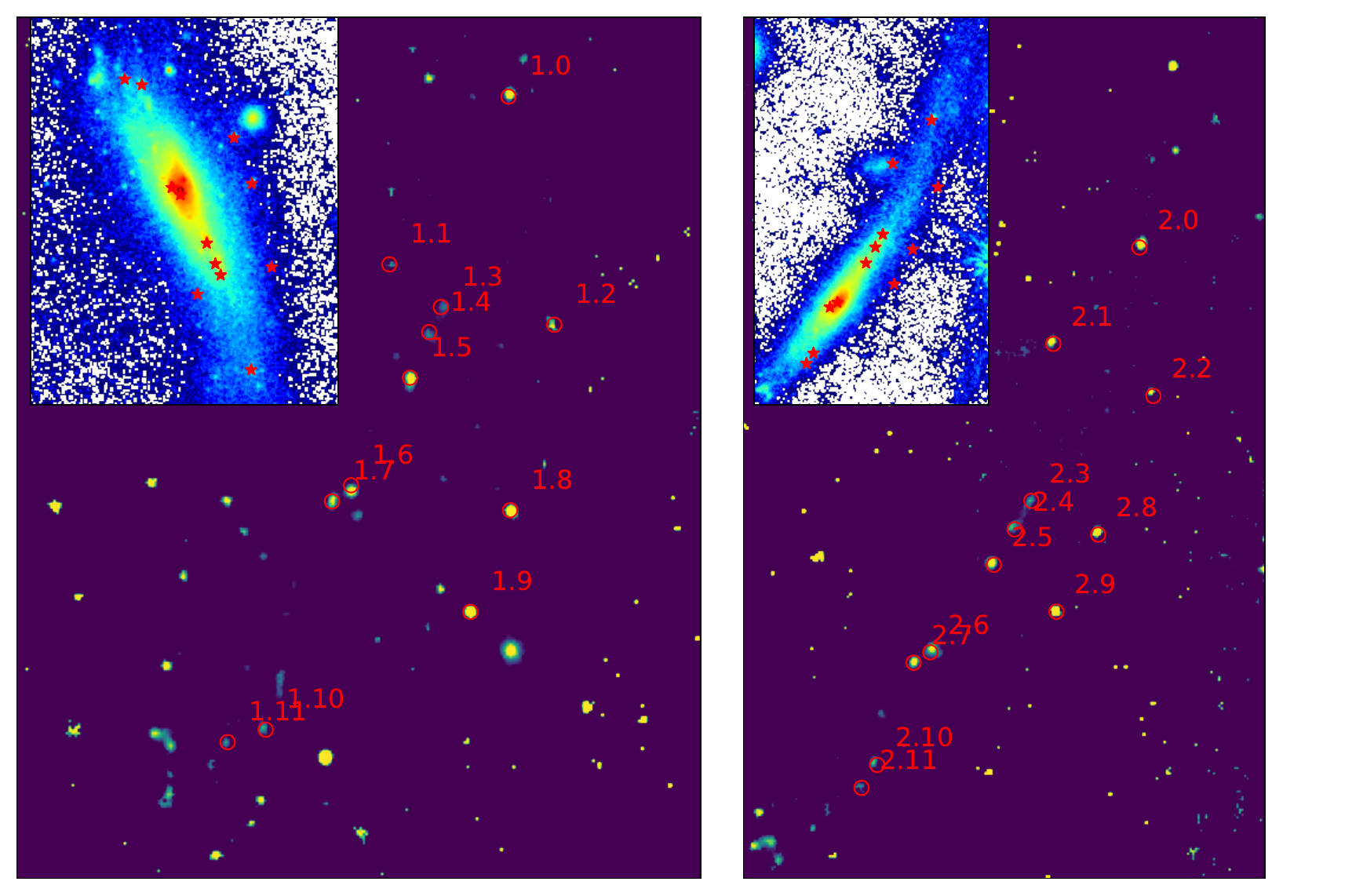}
    \caption{The pairing identification in the SMACS0723 lens arc. This is done on the point source maps generated using the procedure outlined in Appendix~\ref{app: point-sources-locations}. As shown on the point source map, the extended lens arc are removed, leaving minimal amount of noise and a few significant detection of point sources. The original data is shown in the top left inset plots, with the corresponding location of those point sources identified. As a side note, the point source map for arc L1 (on the \textbf{left} panel) is parity flipped and rotated, so as to align with the direction of arc L2 (on the \textbf{right} panel) for easier pairing identification.}
    \label{fig: pairing-map}
\end{figure*}
\begin{figure}
    \centering
    \includegraphics[width=.48\textwidth]{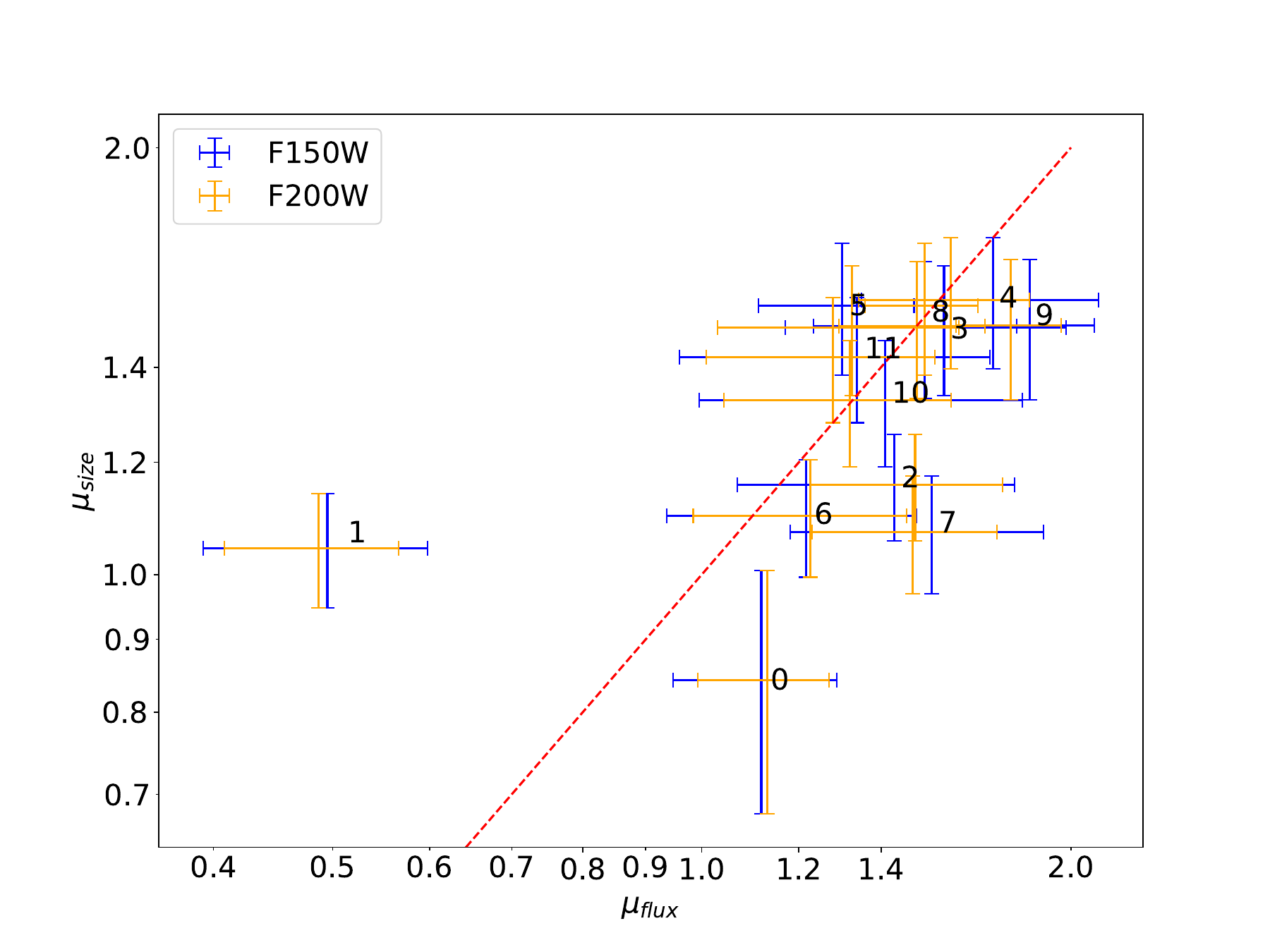}
    \caption{Comparison of the measured flux ratio $\mu_{\rm flux}$ with the inferred angular magnification $\mu_{\rm size}$. The pairing ID of each data point, defined in Fig.~\ref{fig: pairing-map}, is shown right next to the reported error bar. Note that the axes spacing is uniform in the log scale - which is a natural choice when comparing ratios. It is worth mentioning that the measured flux ratios $\mu_{\rm flux}$ is consistent across filters - which is well expected if the point source of concern is indeed lensed by an achromatic gravitational lens.}
    \label{fig: smacs0723-flux-vs-angular}
\end{figure}
\subsection{Measurements}

SMACS~J0723.3--7327 (hereafter SMACS0723) is well known as the first galaxy cluster image released by JWST \citep{smacs0723-lens-model-suyu}. 
At a redshift of $z_l = 0.39$, SMACS0723 contains a total mass of $\sim 10^{15} M_\odot$, 
and thus serves as an ideal strong gravitational lens, deflecting dozens of background galaxies into multiply imaged systems. 
Among these images, several contain multiple ``knots'', which are presumably individual compact star-forming regions in the background galaxies being lensed \citep{mowla_2022}. 

In this paper, we focus our efforts on a particular pair of lens arcs, highlighted in Fig.~\ref{fig: jwst-lens-field}. 
This system has been identified as a legitimate multiply-lensed system in previous cluster-lensing studies \citep{smacs0723-lens-model-dan,smacs0723-lens-model-guillaume,smacs0723-lens-model-suyu,smcas0723-lens-model-pascale,alex-finder-map-smacs0723}. 
Spectrometric follow-up refined the redshift of the source galaxy to $z_s = 2.2$, consistent with the geometrical configuration inferred from lens-modelling analysis \citep{alex-finder-map-smacs0723}.

Of particular interest, this system features clear detections of point-alike sources located around the lens arcs, presumably associated with the source galaxy, forming a pair of lensed ``constellations''. 
Most of the point sources within the lensed constellations are visible in the F150W and F200W filters, while their detectability in other passbands varies from pair to pair. 
As a proof-of-concept study, we thus focus exclusively on F150W and F200W in the following analysis.

Following the image-processing procedures outlined in Appendix~A2, we obtain point-source maps for each of the lens arcs, which define the constellations used in this work. 
The next step is to pair the point sources between the two lens arcs. 
For this proof-of-concept, this pairing is performed manually, based on their relative positions along morphological features of the lens arcs. 
Certainly, this process can be improved by comparing the similarity of the spectral energy distributions (SEDs) of the point sources, although this requires additional steps to remove contamination from the extended arc. 
We leave such improvements for future work. 
The pairings used in this analysis are shown in Fig.~\ref{fig: pairing-map}.

To extract photometry of the point sources, we perform point-spread-function (PSF) matched fitting using model PSFs generated with \textsc{WebbPSF} \citep{perrin_2015} on the latest calibrated JWST images produced by the \textsc{Grizli} pipeline \citep{brammer_2021}. 
We adopt the initial positions from our point-source detection algorithm (Appendix~A2), then refine them via PSF fitting. 
The best-fit PSF centroids are therefore used in the astrometric analysis to estimate $\mu_{\rm size}$.

Because the positions of the point sources are recovered with uncertainties comparable to the pixel scale, these positional uncertainties can noticeably influence the inferred size magnification. 
To account for this, we adopt a conservative positional uncertainty of $\sigma_{\rm pos} = 0.06$~arcsec (the approximate JWST PSF width). 
These uncertainties are propagated into the uncertainties in $\mu_{\rm size}$ via bootstrap resampling.

\subsubsection{Checking for Geometrical Duality}

We apply our algorithm to infer the size magnification $\tilde{\mu}_{\rm size}$ for each identified pair. 
The inferred values are compared against the measured flux magnification $\tilde{\mu}_{\rm flux}$, as shown in Fig.~\ref{fig: smacs0723-flux-vs-angular}. 
The pairing ID (labelled in Fig.~\ref{fig: pairing-map}) is displayed next to each point.

It is evident that most point-source pairs satisfy geometrical duality (i.e. lie near the red dashed line) within error bars. 
First, this confirms that our algorithm reproduces the observed flux magnifications using only astrometric information. 
Second, the measured flux ratios are consistent between F150W and F200W, as expected if each pair consists of multiple images of the same underlying source.

On the other hand, the lack of significant violations of geometrical duality—apart from the lone outlier, pair~1—suggests that no DM sub-structure is detected in this region of the lens. 
In particular, the abundance of sub-haloes cannot be too high, otherwise violations at the level accessible to JWST astrometry and photometry would be expected.

The peculiar behaviour of pair~1 is discussed separately in Section~\ref{sect: discussion}.

\subsection{Evaluating the Sub-structures Sensitivity of This Lens Arc}
\begin{table*}
    \centering
    \begin{tabularx}{\textwidth}{|X|X|X|X|X|}
        \hline
         \textbf{Model} & \textbf{Sources} & \textbf{Main Halo} &  \mbox{\textbf{Mesoscale Perturbers}} & \textbf{Subhaloes}\\
         \hline\hline
         Main cluster halo & \multirow{5}{2.5cm}{16 Point Sources, random locations$^\dagger$ sampled from $\mathcal{N}([0.69,6.99], 0.06) $} & \multirow{5}{2.5cm}{Elliptical tNFW, $M=5\times10^{14} M_\odot$, $c=0.26$, axis\_ratio=$0.85$} & - & \multirow{2}{*}{-}  \\
         \cline{1-1}\cline{4-4}
         + cluster members &  & & \multirow{4}{2.5cm}{two tNFW,
         $(M,c)= (2\times 10^{12}, 9.7)$ \& $(10^{11},13.5)$} &  \\
         \cline{1-1}\cline{5-5}
         + Massive subhalo & &  & & tNFW,$(M,c) = (10^{9}, 22.2)$  \\
         \cline{1-1}\cline{5-5}
         + Small subhalo & &  & & tNFW,$(M,c)= (10^{8}, 28.6)$ \\
         \cline{1-1}\cline{5-5}
         + Over-concentrated small subhalo & &  & & tNFW,$(M,c)= (10^{8}, 100)$  \\
         \hline
    \end{tabularx}
    $^\dagger$ Gaussian distributed, the centre is defined in the source plane, with the origin of the coordinate system centred at the lens centre.
    \caption{Summary of the lens model parameters used for creating systems similar to the SMACS0723 lens arc. All the positions reported above are in the unit of arcsec, masses in unit of $M_\odot$, and concentration parameters $c$ are dimensionless. tNFW model is the truncated NFW profile proposed in \citep{tnfw-formalism-oguri} where the truncation radius is chosen to be $30 r_{\rm vir}/c $ with $r_{\rm vir}$ being the Virial radius.}
    \label{tab: analog-smacs0723-details}
\end{table*}

\subsubsection{The Main Lens: Cluster Halo}
\begin{figure*}
    \centering
    \includegraphics[width=.98\textwidth]{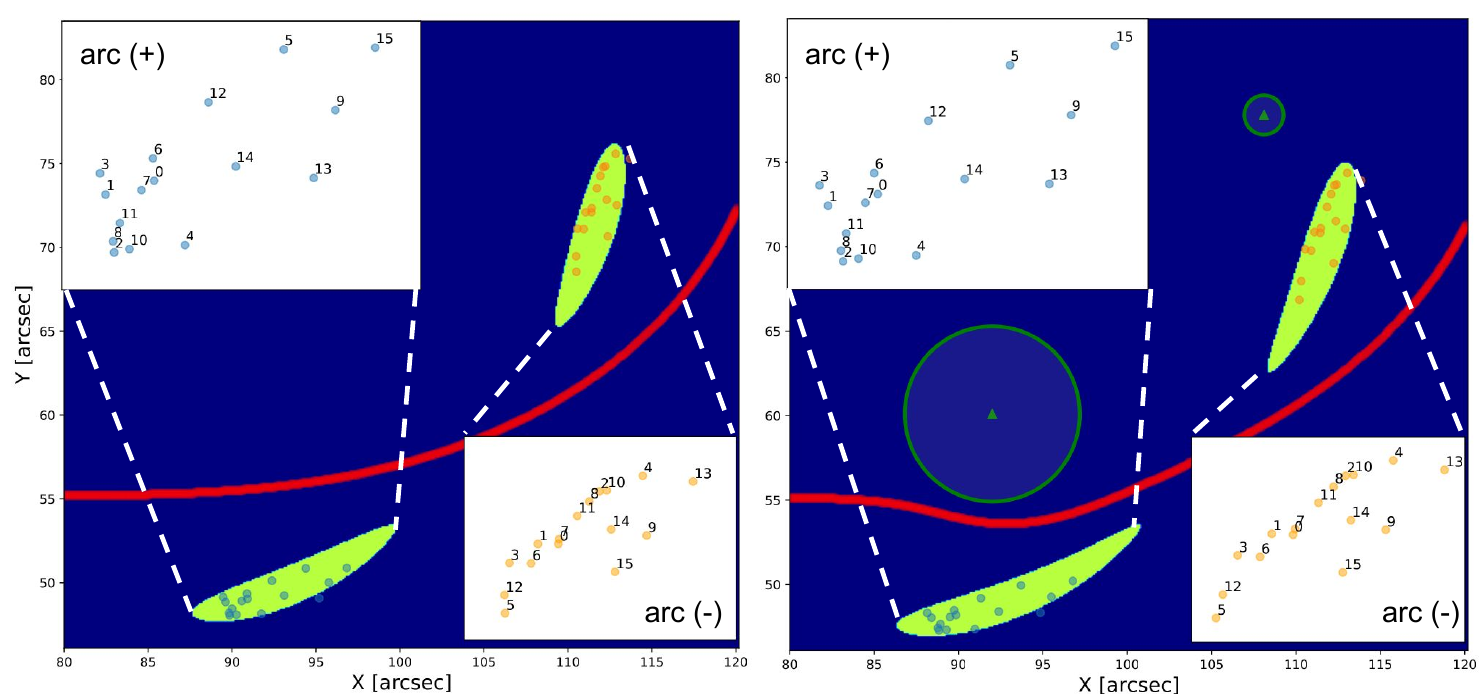}
    \caption{The configuration of the analogous lens arcs system which qualitatively resemble the SMACS0723 arcs studied in this work. On the \textbf{left} panel, we show the macroscopic lens model, with the bold red line bisecting the image being the critical curve. 16 point sources are put in the source plane and their corresponding lensed image locations are shown with more details in the zoom-in plots. For clarity, the X-Y axes in the zoom-in plots are not scaled equally. Each of the lens arcs is marked by $(\pm)$ depending on the image parity. On the \textbf{right} panel, the image configuration is shown when 2 mesoscale perturbers are added in the lens potential. The locations of these perturbers, and their corresponding Einstein radius, are plotted in the triangles enclosed by circles.}
    \label{fig: analogous-jwst-arc}
\end{figure*}
\begin{figure*}
    \begin{subfigure}[b]{0.48\textwidth}
    \centering
    \includegraphics[width=1\textwidth]{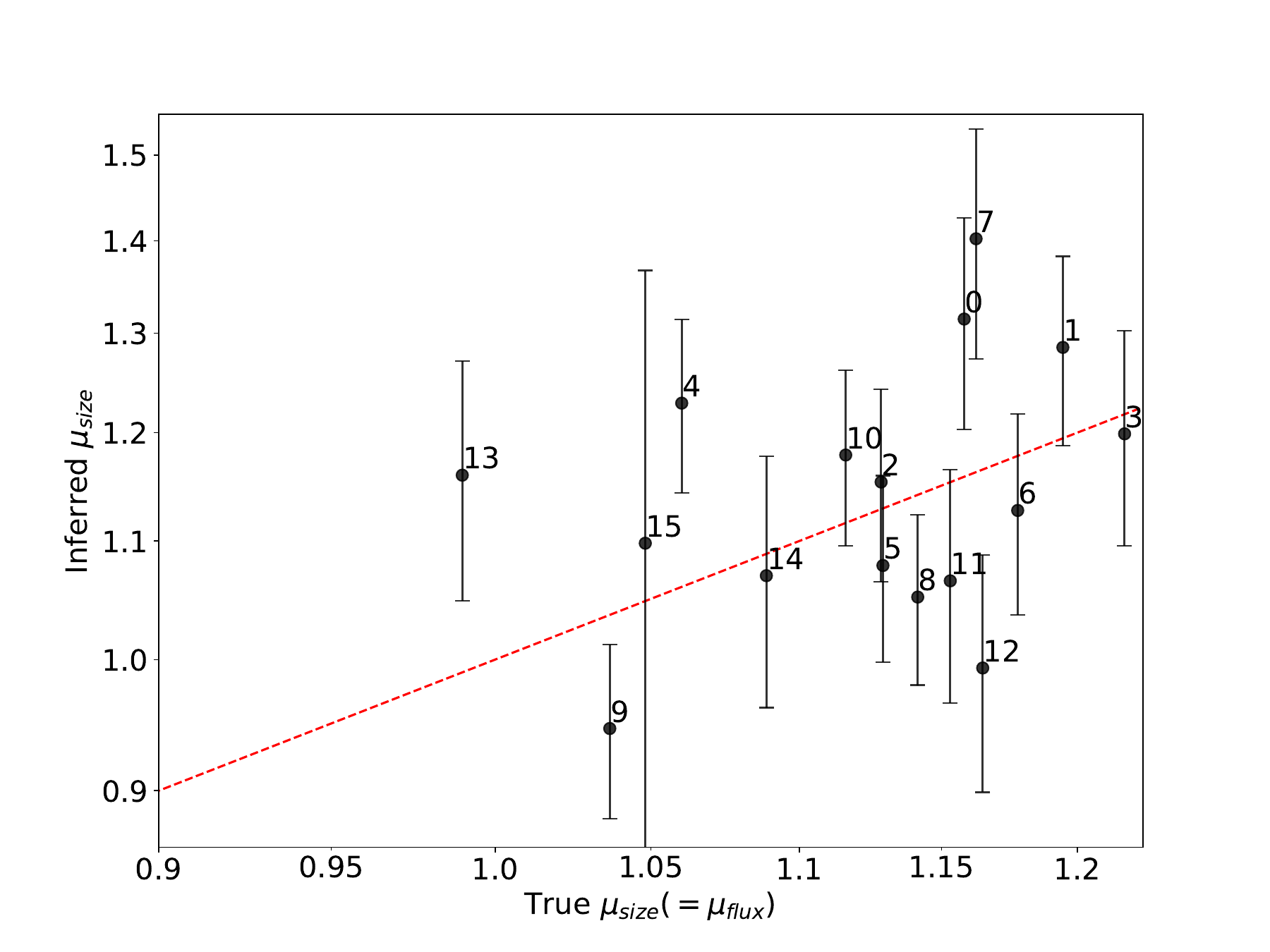}
    \end{subfigure}
    \begin{subfigure}[b]{0.48\textwidth}
    \centering
    \includegraphics[width=1\textwidth]{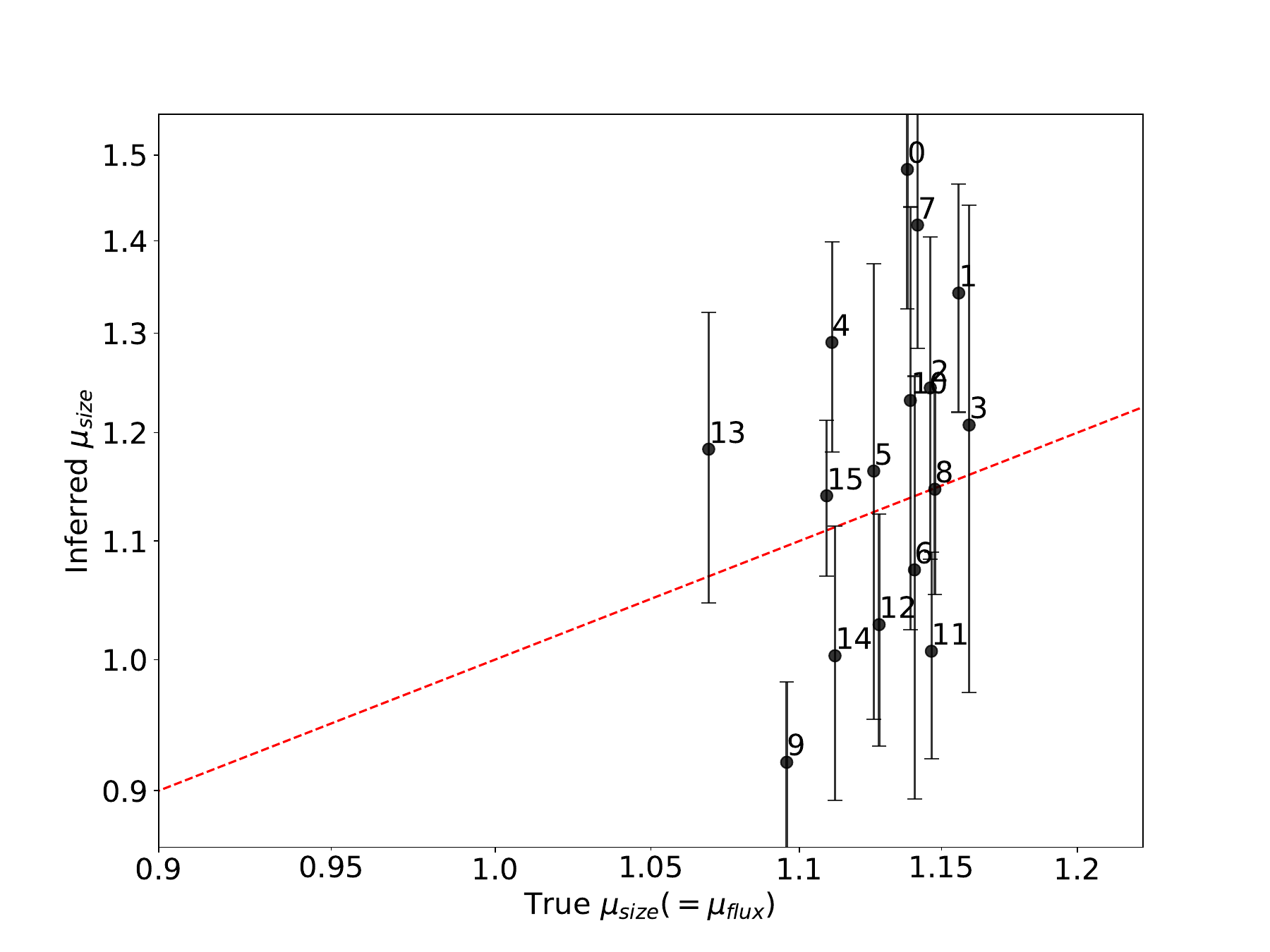}
    \end{subfigure}
    \caption{The underlying magnifications injected in the simulation (`true'), versus the inferred magnification $\mu_{\rm size}$ by applying the \textit{Quad Jacobian} estimator on the multiply lensed constellation. 
    Note that when the flux is measured, it would faithfully reproduce the underlying true magnification, so that the x-axis can also be interpreted as the flux magnification $\mu_{\rm flux}$.
    On the \textbf{left}, we show the simulation case consisting of the smooth macroscopic cluster potential only, while on the \textbf{right}, we show the results when 2 meso-scale perturbers are added to mimic a more realistic cluster environment. One can see that in both of the cases, the \textit{Quad Jacobian} estimator can indeed reasonably predict the underlying magnification. }
    \label{fig: geo-dual-analog-sim}
\end{figure*}

As discussed in the previous subsection, we measured the size magnification versus the flux magnification and found no significant departures from geometrical duality, except for point-source pair~1 — which we analyse separately in Section~\ref{sect: discussion}. 
In this subsection, we perform simulated observations of a system similar to the SMACS0723 lens arcs and assess how sensitive and robust this particular configuration is when our proposed method is applied.

In particular, we inject CDM sub-haloes and cluster members into a fold configuration and assess the expected observational signatures. 
This exercise provides a reference for interpreting the non-detection of any signal in our geometrical duality algorithm, especially for understanding the systematics involved in constructing $\mu_{\rm size}$ estimators.

We first generate a lens system that produces a pair of lens arcs similar to the lens arc in SMACS0723. 
The adopted parameters are summarised in Table~\ref{tab: analog-smacs0723-details}, and we show the visual rendering of the simulated lens arc on the left panel in Fig.~\ref{fig: analogous-jwst-arc}. 
The main cluster halo is an elliptical truncated NFW (tNFW) profile \citep{nfw-profile,tnfw-formalism-oguri} of $M = 5 \times 10^{14}\,M_\odot$ with a concentration of $c = 0.26$, which is slightly below the expected value from the mass--concentration relation \citep{mass-concentration-relation-2007}. 
We randomly place $N = 16$ point sources around the centre of the extended source, with source-plane positions drawn from a Gaussian distribution with standard deviation $\sigma_{\rm ps} = 60$~mas.

We run our size magnification estimation pipeline on the simulated data, and verify whether the obtained size magnifications are consistent with the `ground truth' magnification derived from the simulation parameters. 
Indeed, the true magnification is recovered within a standard deviation of $\sigma_{\delta\mu/\mu} \approx \pm 10\%$. 
The resulting lens image configuration is shown in the inset zoom-in plots in Fig.~\ref{fig: analogous-jwst-arc}. 
It is worth noting that we also show the lens arc in the image configuration only for illustrative purposes. As our algorithm used solely the point sources (‘constellations’) along the lens arcs, the extended light profile of the lens arc itself was not used in the subsequent discussions. We thus ignored the extended light profile of the lens arc in the subsequent discussions.

\subsubsection{Mesoscale Perturbations: Effect of Cluster Members}

We also test the size magnification estimators in the presence of a few neighbouring cluster members that distort the critical curve into more complicated geometry and introduce mesoscale astrometric shifts. 
The injected cluster members are listed in the fourth column of Table~\ref{tab: analog-smacs0723-details}. 
The local configuration of the perturbed critical curves is shown on the right panel in Fig.~\ref{fig: analogous-jwst-arc}. 
The locations of these mesoscale perturbers are marked with green triangles, along with their corresponding Einstein radii.

In terms of the point sources, the average astrometric shifts induced by the injected cluster members are typically $\lesssim 500$~mas. 
As the size estimator uses relative separation between the point sources (i.e. the baselines) to infer the underlying lensing magnification, the measurable change in the baselines from those mesoscale structures, can be captured by the estimator as well. Consequently, we found that our size magnification estimators can still faithfully recover the true underlying magnification, as evident in Fig.~\ref{fig: geo-dual-analog-sim}.

\subsubsection{Dark Matter Sub-haloes} \label{sect: sub-halo-sim}
\begin{figure}[t]
    \centering
    \includegraphics[width=.48\textwidth]{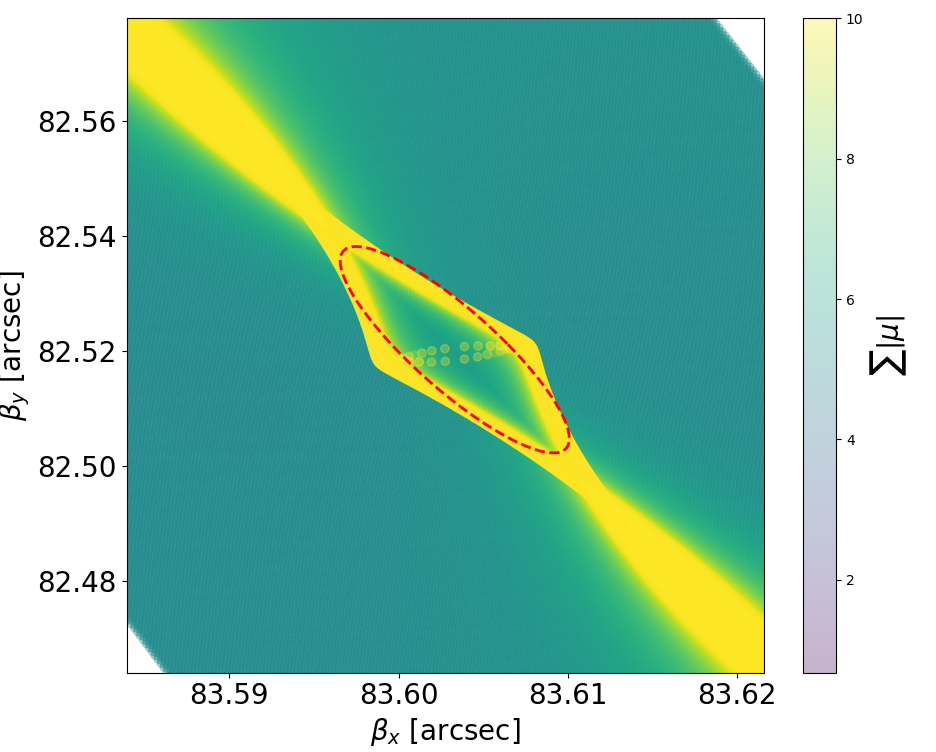}
    \caption{The source plane magnification map for the case with an over-concentrated subhalo. In this case with a super-critical subhalo, the subhalo can introduce further image splitting, and the magnification shown here is the total magnification summed over all the multiply images. The dashed ellipse marks the definition of $r_{\rm eff}=1$, which is an analytical approximation of the caustic size using the Einstein radius in the image plane.}
    \label{fig: r_eff-defintion}
\end{figure}
\begin{figure}
    \centering
    \includegraphics[width=.48\textwidth]{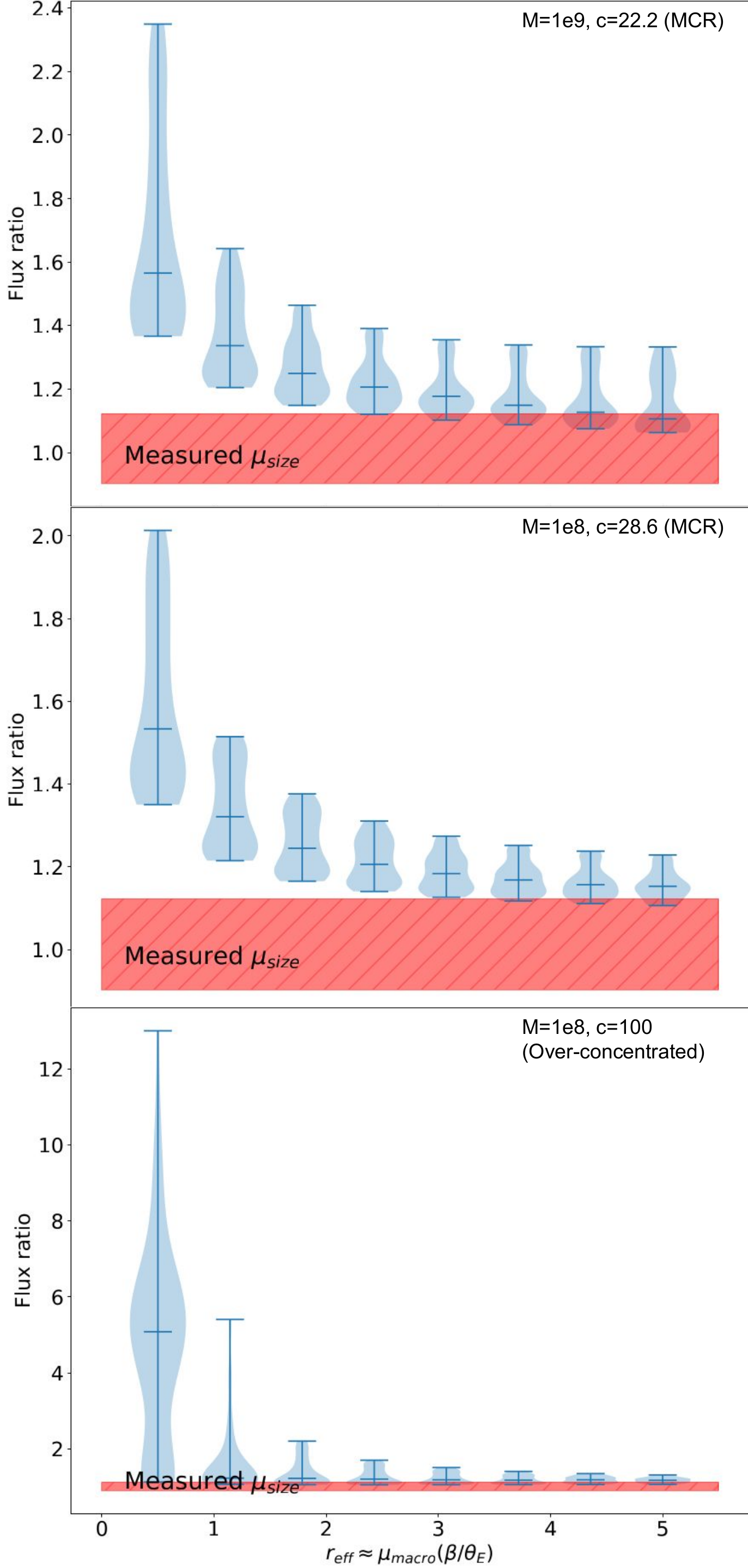}
\caption{The model prediction of flux excesses created by different sub-haloes, as a function of the dimensionless effective separation $r_{\rm eff}$ from the centre of the sub-halo (defined graphically in Fig.~\ref{fig: r_eff-defintion}). As there is angular dependence in the flux excess, we show the distribution of flux excess along all the angular directions at the same $r_{\rm eff}$. For reference, we also show the magnification ratio $\mu_{\rm size}$ inferred using \textit{Quad Jacobian} estimator with the horizontal hatched band. From top to bottom: massive sub-halo, small sub-halo and over-concentrated small sub-halo as defined in Table ~\ref{tab: analog-smacs0723-details}.}
    \label{fig: flux-excess-model}
\end{figure}

\begin{figure*}
    \centering
    \includegraphics[width=.95\textwidth]{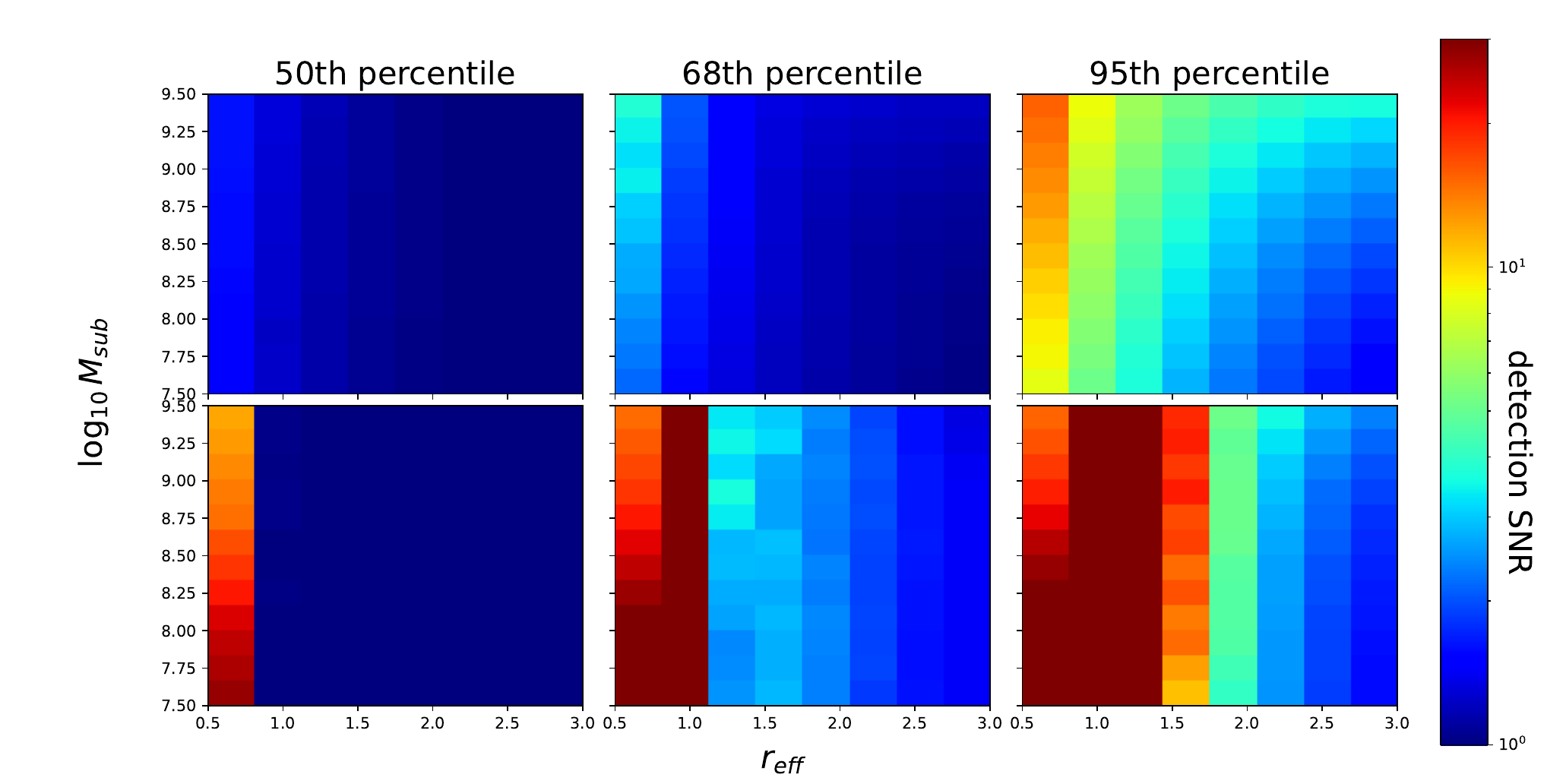}
\caption{The detection signal-to-noise ratio (SNR) of the simulated NFW haloes of different masses (y-axis) located at various distance in the image plane away from a point source (x-axis). The distance quoted here are the dimensionless distance normalised to the Einstein radius of the corresponding halo mass. \textbf{First row:} assuming the DM sub-haloes follow from the CDM mass concentration relation; \textbf{Second row:} assuming over-concentrated haloes with concentration artificially made 5 times higher than the CDM expectation. 
As the level of detectable flux excess depends on the exact 2D location of the sub-haloes, instead of the 1D straight line distance, here we consider the characteristic values of the detection SNR distribution. 
The three columns show different summary statistics (percentiles) of the underlying detection SNR distribution. }
    \label{fig: sensitivity-map}
\end{figure*}

Finally, we study the effect of sub-haloes in this configuration. 
This allows comparison of our measurements with expectations from model-based sub-halo finders. As a proof of concept, we test 3 limiting cases of sub-haloes. 
The injected sub-haloes are fixed at masses of either $M_{\rm sub} = 10^8\,M_\odot$ or $10^9\,M_\odot$ characterised by the truncated NFW profile without ellipticity, each with different concentration. 
In the first case, the concentration of the sub-haloes is fixed to align with the standard mass--concentration relation \citep{mass-concentration-relation-2007}, taking the value of $c_{\rm sub} \sim 20$. 
Alternatively, in the second case, we consider an over-concentrated sub-halo, taking a rather extreme value of $c_{\rm sub} = 100$. 
This is motivated by the fact that a considerable portion of the claimed detections using `gravitational imaging' \citep{overconcentrated-halo-minor2021} 
infer very high concentrations (up to $c_{200}\!\sim\!1500$). 
The parameters used are summarised in Table~\ref{tab: analog-smacs0723-details}.

Instead of injecting these sub-haloes into a fixed location, we evaluate the sensitivity — namely, the furthest separation from the point sources for which the sub-haloes still remain detectable. 
To generalise the results, we present the distance $r_{\rm eff}$ in dimensionless units, which captures the effect of both the background cluster potential and the sub-halo mass. 
Consequently, $r_{\rm eff}$ is defined as the physical separation to the sub-halo normalised to the Einstein radius of the sub-halo in isolation, $\theta^{\rm sub}_E$, and the lensing deformation tensor of the background cluster, $\partial\beta_i/\partial\theta_j\big|_{\rm cluster}$. 
The deformation effect is typically anisotropic, squeezing the circular Einstein ring into a unit ellipse. 
This unit ellipse is visualised in Fig.~\ref{fig: r_eff-defintion} as a dashed line on top of the magnification map. 
It approximately follows the caustics, which is itself an iso-magnification contour, making it a useful definition of the sub-halo’s region of influence.
The parameter $r_{\rm eff}$ then parametrises the concentric enlargement (or contraction) of the unit ellipse.

In the case of the over-concentrated sub-halo, an extra sub-critical curve (and thus caustics) emerges. Crossing the caustics would result in further image splitting, creating two extra copies of the point source. As the image splitting is not resolvable by the diffraction-limited telescope, one could only see the total flux contributed by such image splitting configuration, thus generating extreme flux magnification. 
This is confirmed in our sub-halo injection pipeline: in Fig.~\ref{fig: sensitivity-map}, we show the flux magnification at different $r_{\rm eff}$. 
One can see that when $r_{\rm eff} < 1$ (representing the case that the point source is located inside the sub-halo’s caustics), the flux magnification is maximised. 
Because there is residual angular dependence in the magnification, we plot the distribution of the flux along different angular positions at the same $r_{\rm eff}$ in the violin plot.
In general, the chance of observing excess flux magnification is maximised at $r_{\rm eff} < 1$, which defines the sensitivity area of our sub-structure finding algorithm.

As a final check for establishing detectability, we chart the sub-halo detection signal-to-noise ratio (SNR) at different $r_{\rm eff}$ as a function of sub-halo mass. 
The detection `signal' is defined as the observed flux magnification over the measured size magnification \textbf{(i.e. $\mu_{\rm flux}/\mu_{\rm size}$)}. 
This signal, should equal unity in the absence of sub-halo lensing. 
As for the `noise', it is defined by the measurement uncertainty from astrometric error in the measured size magnification. 
The SNR is thus the significance of the deviation from unity. 
The results are visualised in Fig.~\ref{fig: flux-excess-model}. 
As the flux magnification varies across the same ellipse, we examine summary statistics of the detection-SNR distribution — in particular, the $50\%$, $68\%$, and $95\%$ percentiles.

When the sub-haloes follow the MCR (top row), and are therefore more diffuse, an extra critical curve does not emerge. 
In this case, the detection SNR is modest. 
Only in a handful of locations near the sub-halo centre does the SNR become large enough for statistically significant detection. 
By contrast, when the sub-haloes are over-concentrated, the detection SNR is substantially boosted: the mean SNR often exceeds $\sim 10$ inside the unit ellipse; the 95th percentile reaches $\sim 50$ around the unit ellipse, and the SNR remains $>5$ out to $r_{\rm eff} \approx 2$.

\subsection{Interpretations of the Observed Duality}
\label{sect: discussion}
Evidently from Fig.~\ref{fig: smacs0723-flux-vs-angular}, the point source pair 1 reveals a significant discrepancy from geometrical duality: the measured flux ratio is $\mu_{\rm flux}\sim 0.5$ (versus $\mu_{\rm size} = 1.04 \pm 0.09$), suggesting that its counter image on arc 2 is brighter. 
This is highly unexpected, as both the size magnification $\mu_{\rm size}$ (estimated via \textit{Quad Jacobian} estimator) and the flux ratios of other nearby point sources suggest a brighter arc 1. 
If such a detection is legit, the results we obtained in Section~\ref{sect: sub-halo-sim} indicate that the anomalous flux ratio can be explained by a sub-halo located within a few Einstein radii of point source 1 on arc 2, thereby enhancing its flux beyond theoretical expectations. 
Furthermore, as the anomaly is localised at point source 1 instead of affecting other nearby point sources, it cannot be caused by a massive sub-halo with an Einstein radius larger than the separation between point source 1 and its neighbours.

Non-lensing interpretations, however, are viable possibilities. 
From the visual appearance of point source 1 on arc 2 (labelled as 2.1), as shown in the inset plot on the right panel of Fig.~\ref{fig: pairing-map}, this particular point source is overlaid on an extended diffuse emission region. Furthermore, such an extended feature is absent in its counter image. 
Even if point source 1 is indeed a lensed pair, this extended feature could be a foreground object coincidentally situated along the line of sight. 
The under-subtraction of the extended feature when performing PSF matching photometry may be a reason for overestimating the flux of point source 2.1, making it appear brighter on arc 2. 
Alternatively, the point source pair 1 may not be multiple images of the same object. 
That their locations appear consistent with a lensing interpretation could be a pure coincidence. 
The limited photometric information---specifically that the system is only detected in F150W and F200W---makes it difficult to exclude such a possibility. 
However, this would also require an extra coincidence in their colours, such that the observed consistency of the flux ratios across two different filters can be explained.

We do not attempt to assert the nature of the point source pair 1 in this work. 
Further observations that constrain the spectral energy distribution (SED) of this system will be useful in verifying whether it comprises two unrelated objects by chance, or confirming that it is a genuine lensed pair. 
In the case that the point source pair possesses a highly similar SED profile, further follow-up analysis with parametric lens modelling of this system may better quantify the nature of the anomalous flux ratio.

\subsection{Subhaloes Abundance Constraints}
Finally, it is useful to estimate the equivalent constraint on the abundance of CDM sub-haloes. 
Here, we present a very rough estimate; a detailed analysis of the underlying systematics and selection biases is required to derive a more accurate result, which is beyond the scope of this work.
From Fig.~\ref{fig: flux-excess-model}, it is shown that only point sources located within a few Einstein radii $<\zeta \theta_E$ of the sub-haloes can generate the necessary excess flux to be identified by our algorithm. 
In this analysis, we choose $\zeta=3$; but in general, with more point sources and more accurate measurements of $\mu_{\rm size}$, the detection threshold can be improved by reducing the uncertainty, thereby allowing for a tighter $\zeta$.

If the surface density of sub-haloes is $\Sigma_{\rm sub}$, the chance of detecting excess flux from a single point source is $p_{\rm det} = \Sigma_{\rm sub} \cdot \pi (\zeta\theta_E)^2$.
This is a binomial experiment with the chance of success equal to $p_{\rm det}$.
If the observed anomaly of the point source pair 1 is \textit{not} a lensing event, the non-detection across a pair of lens arcs, each with 12 point sources, translates into an upper bound of $p_{\rm det} < (1-(\Delta/2)^{1/(2 \times 12)}) = 0.142$ within a $95\%$ confidence interval (i.e., $\Delta = 0.05$).
If we pick the representative value of $M_{\rm sub}=10^8 M_\odot$, which corresponds to $\theta_E = 0.0232$ arcsec $= 0.388$ kpc, this translates to:
\begin{equation}
\Sigma_{\rm sub}^{\rm (this\, work)} < 0.300 \, \text{kpc}^{-2}.   \, \text{(95\% credible interval)} 
\end{equation}
The CDM expectation for $\Sigma_{\rm sub}$ depends on many factors. 
It has been shown that the shape of the sub-halo mass function is almost scale invariant, with $dN/d(\log M_{\rm sub}) \propto M_{\rm sub}^{-1}$ \citep{cdm-cluster-subhalo-sim}, but the normalisation scales with the host halo mass $M_{\rm main}$. 
Furthermore, the sub-halo mass fraction for different $M_{\rm sub}$ varies as a function of radial distance from the cluster centre, which is expected as a consequence of mass segregation. 
To get a rough sense of the order of magnitude, we quote the value from an analysis of quasar flux anomalies, which claimed that CDM simulations of $M_{\rm main} = 10^{13} M_\odot$ suggest \citep{flux-anomalies-vs-cdm-sim-2015}:
\begin{equation}
    \Sigma^{\rm (CDM)}_{\rm sub} \sim 10^{-2} \, {(h^{-1} \text{kpc})^{-2}},
\end{equation}
as the surface number density of $M_{\rm sub}=10^8 M_\odot$ sub-haloes around the critical curve defined by the main lens.
We emphasise that the number quoted here is only for illustrative purposes, as actual CDM predictions could be more complex, such as the dynamical status of the cluster which governs the tidal disruption of small sub-haloes.
Note that in our observed system, the mass of SMACS0723 is of the order $M_{\rm main} = 10^{15} M_\odot$, so it is reasonable to expect that the actual CDM prediction applied to SMACS0723 would fall within $10^{-3} - 10^{-1}$ kpc$^{-2}$, which is consistent with our observations.

\section{Discussion and Conclusion} \label{sect: conclusion}
Unlike the more traditional methods for detecting dark matter sub-haloes via parametric modelling, which are subject to degeneracies and convergence issues arising from the huge latent parameter space, in this work we propose a fully model-independent method to detect DM sub-structures, complementing  parametric approaches. 
We have exploited the Optical Liouville Theorem to identify locations where small-scale compact gravitational structures reside.
These small-scale structures can be sub-haloes having corresponding Einstein radii smaller than the diffraction limit of the telescopes, thus producing negligible astrometric perturbations yet potentially sizeable flux anomalies. 
This is achieved by comparing the observed size magnification $\mu_{\rm size}$ and the flux magnification $\mu_{\rm flux}$, which should be equivalent to each other as a consequence of the Optical Liouville Theorem --- a relation that we term the geometrical duality. 
Constructing estimators for the size magnification from observed data can be done without building any explicit lens model or source model.
We have shown that this is possible in strongly-lensed `constellations' of compact star clusters appearing as point-like sources within large, lensed galaxy images.
We have applied this method to a specific pair of lensed constellations identified in the lensing cluster SMACS0723 in deep JWST images. We have found that most of the point source pairs in this system indeed follow the geometrical duality, with the exception of one pair (pair 1).
We then performed simulations of such constellations using configurations analogous to the observed case in SMACS0723, finding that the presence of sub-haloes can generate the observed violation of the geometrical duality. While this is consistent with the sub-halo lensing interpretation, we cannot rule out other explanations given the current data, which we aim to investigate further.

The systematics of this method have also been evaluated using our simulations. We have found that the dominant source of systematics is the precision with which we can determine $\mu_{\rm size}$. 
The performance of our proposed estimator degrades significantly when the lensed image of the constellation is highly stretched along one direction, making the locations of the lensed point sources appear nearly collinear. 
This effectively reduces the system to one dimension from which we must infer the two-dimensional size magnification. 
This systematic is related to the cluster lens--source alignment because, for fold and cusp configurations, the magnification is highly directional.

Recent JWST observations of `exotic' cluster lens configurations (for a review see \citet{atlas-exotic-lens}) --- the umbilic configuration \citep{meena-hu-lens-paper1,meena-hu-lens-paper2,meena-hu-lens-paper3,meena-hu-lens-paper4,meena-hu-lens-theory-analysis} in which the tangential critical curve and the radial critical curve almost intersect --- offer the desired setup with more isotropic magnification. 
We expect that with such systems, the precision in determining $\mu_{\rm size}$ can be substantially improved, making them ideal sub-structure detectors. 
These umbilic lenses were observed in archival data from the Hubble Space Telescope and were recognised as a sensitive DM sub-halo detector (using the ratio of signed sum of image magnification) \citep{lagattuta-hu-lens-dm-constraint}. 
A follow-up analysis performed by the same team \citep{david-rxj0437-substructure} demonstrated a viable substructure candidate using the flux ratio argument applied on a strongly lensed `constellation'. 
In fact, a dedicated observation programme with JWST had been carried out recently (\textit{GO Cycle 3, Proposal 6207; PI: D. Lagattuta}), targeting the cluster lens RXJ0437+00.
Further analysis of this umbilic lens will be performed soon, providing much improved constraints on the abundance of small sub-haloes.

It is also important to better quantify the interplay between linear astrometric shifts and non-linear magnification in the context of sub-halo lensing, as the essence of our method is to exploit the disagreement between $\mu_{\rm size}$ inferred from astrometry and $\mu_{\rm flux}$ inferred from flux magnification.
Developing a better understanding of this effect will be crucial to clarify the sensitivity of our proposed geometrical duality method to dark matter sub-haloes of different masses. 
While simulated observations can address this issue to some extent, (semi-)analytical modelling would allow for a deeper insight into the scalability of the geometrical duality method, and better generalisation to a wider sub-halo mass range; this is now being investigated by (Fung et al. \textit{in prep.}).

This work is just a beginning towards a better understanding of DM sub-structures. 
With more lens data from the ongoing JWST surveys, this method can provide complementary constraints on DM sub-structures soon.

\section*{acknowledgments}
The authors acknowledge the helpful feedback from Jose M. Diego, Wolfgang Enzi, Carlos Frenk, Richard Massey, Keiichi Umetsu and Tao Liu.
L.W.H. Fung acknowledges the useful discussions with Albert W.K. Lau, Alex Chow and Jiashuo Zhang.
Fung performed the majority of this work when he was in the Hong Kong University of Science and Technology, before he moved to Durham University, where he is then supported by the UK Science and Technology Facilities Council via grant ST/X001075/1.
Fung is a Beloe Scholar and acknowledges the partial financial assistance by the \textit{Worshipful Company of Scientific Instrument Makers}. 
S.K. Li and J. Lim acknowledge support from the Research
Grants Council (RGC) of Hong Kong through the General Research
Fund (GRF) 17312122.
S.K. Li acknowledges the generous support from the Croucher Foundation through the Croucher Postdoctoral Fellowship.

\appendix
\section{Pipeline for simulating lens images}
\subsection{Ray tracer for solving lens equation}
To perform the simulations, we use our own bespoke ray-tracing code, built on top of \textsc{jax} which utilises automatic differentiation \citep{jax-software} to obtain analytical derivatives. 
As shown in Section~\ref{sect: classic-optics-review}, the lensing deflection and magnification are calculated by taking derivatives of the lens potential. 
This is crucial for our application: the masses of the lensing deflectors span a vast dynamical range, from sub-haloes of $10^8 M_\odot$ to galaxy clusters of $10^{15} M_\odot$, with $\sqrt{(15-8)} = 3.5$ orders of difference in the amplitude of the implied deflection field. 
Furthermore, lensing magnification is highly non-linear in the strong-lensing regime, and for point sources, can reach arbitrarily high values. 
The multi-scale nature of this lensing problem indicates that the traditional approach of discretising the image plane would not be accurate enough for our purposes. 
This automatic-differentiation-based approach for performing numerical ray tracing therefore provides us with better estimates of the lens observables. 
We use this tracer to densely sample the image plane along the lens arc, generating grid points to refine the positions of the point sources in the subsequent step.

\subsection{Locating point sources in the image plane} \label{app: point-sources-locations}
While tracing the image-plane location of an extended source (e.g. a galaxy) can be easily done using a ray-tracing search over the source grids, 
this is not sufficiently precise to trace a point source to sub-grid accuracy. 
Here, we take a hybrid approach to carefully interpolate the image-plane location of a specific point source using a piecewise linear mapping. 
By inverting this piecewise linear mapping, we consequently
obtain a mapping from the source-plane coordinates
directly to the image plane.

To obtain a pair of grid points for interpolation, we first perform ray tracing from the image-plane grid to the source-plane positions. 
The source-plane points are not regularly spaced, highlighting the non-linear nature of strong gravitational lensing. 
The generation of grid pairs is done separately for 
regions separated by the critical curve, to avoid singularities in the mapping. 
As we carefully separate the sets of source- and image-plane grids by the critical curve, it can be guaranteed that there is only one lensed image in each pair of grids, ensuring a one-to-one mapping for each pair. 
Two-dimensional interpolation is then performed by utilising Delaunay triangulation to obtain a piecewise bilinear mapping from the source plane back to the image plane, thereby effectively reversing the direction of the ray tracing.

\section{Assessment of the Performance of Size Magnification Estimators} \label{app: estimator-performance}
\begin{figure*}
    \centering
    \includegraphics[width=0.95\linewidth]{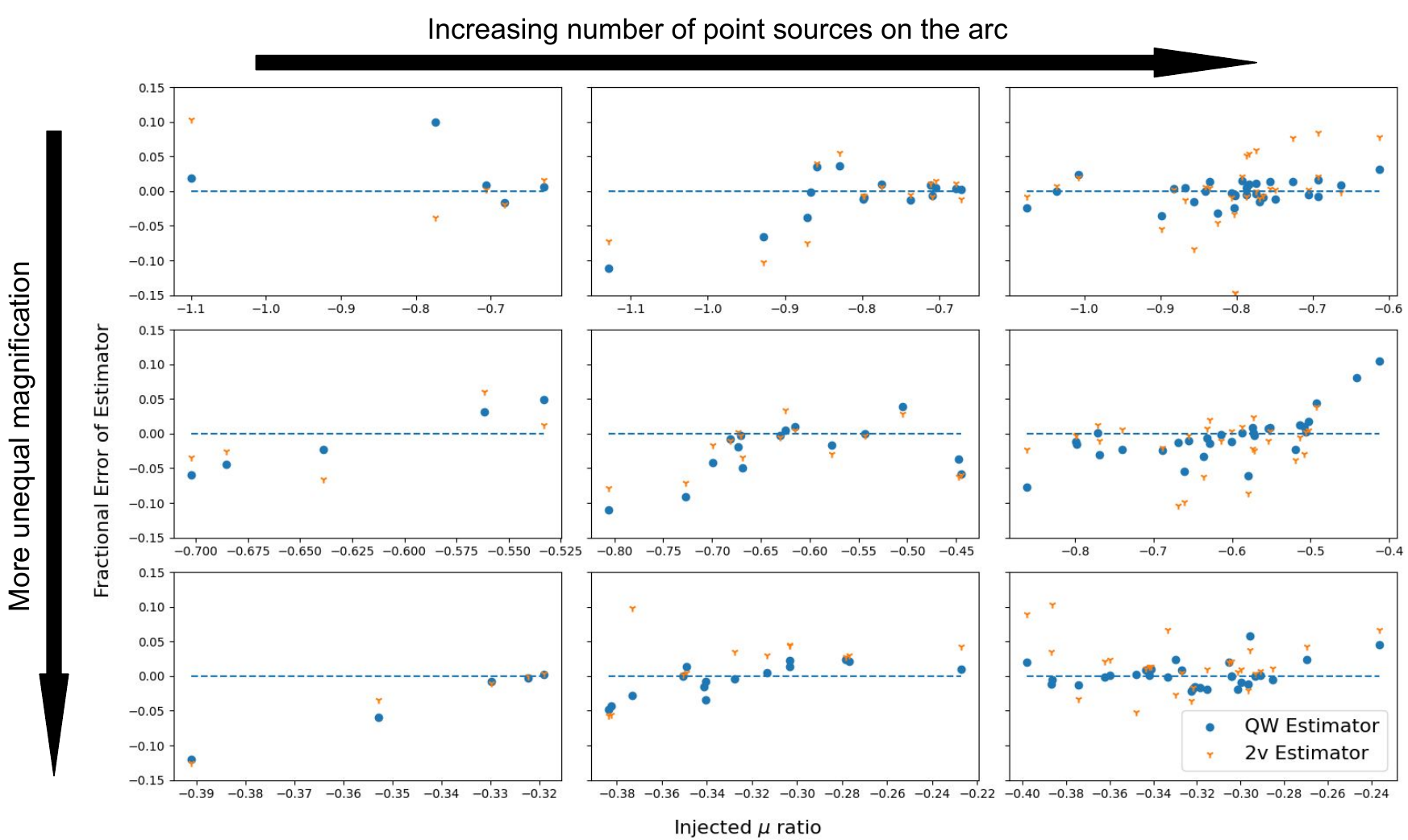}
    \caption{The fractional estimation error of each of the estimators. We ran 9 sets of simulation, going from the left to the right panels, we increase the number of point sources that constitute the `constellation' in the source plane gradually. Going from the top to the bottom panels, we make the relative magnification among the 2 lens arcs more extreme. 
    On the x-axis, we showed the injected magnification ratio used in the simulation. 
    }
    \label{fig: estimators-performance-grid}
\end{figure*}
\begin{figure}
    \centering
    \includegraphics[width=0.95\linewidth]{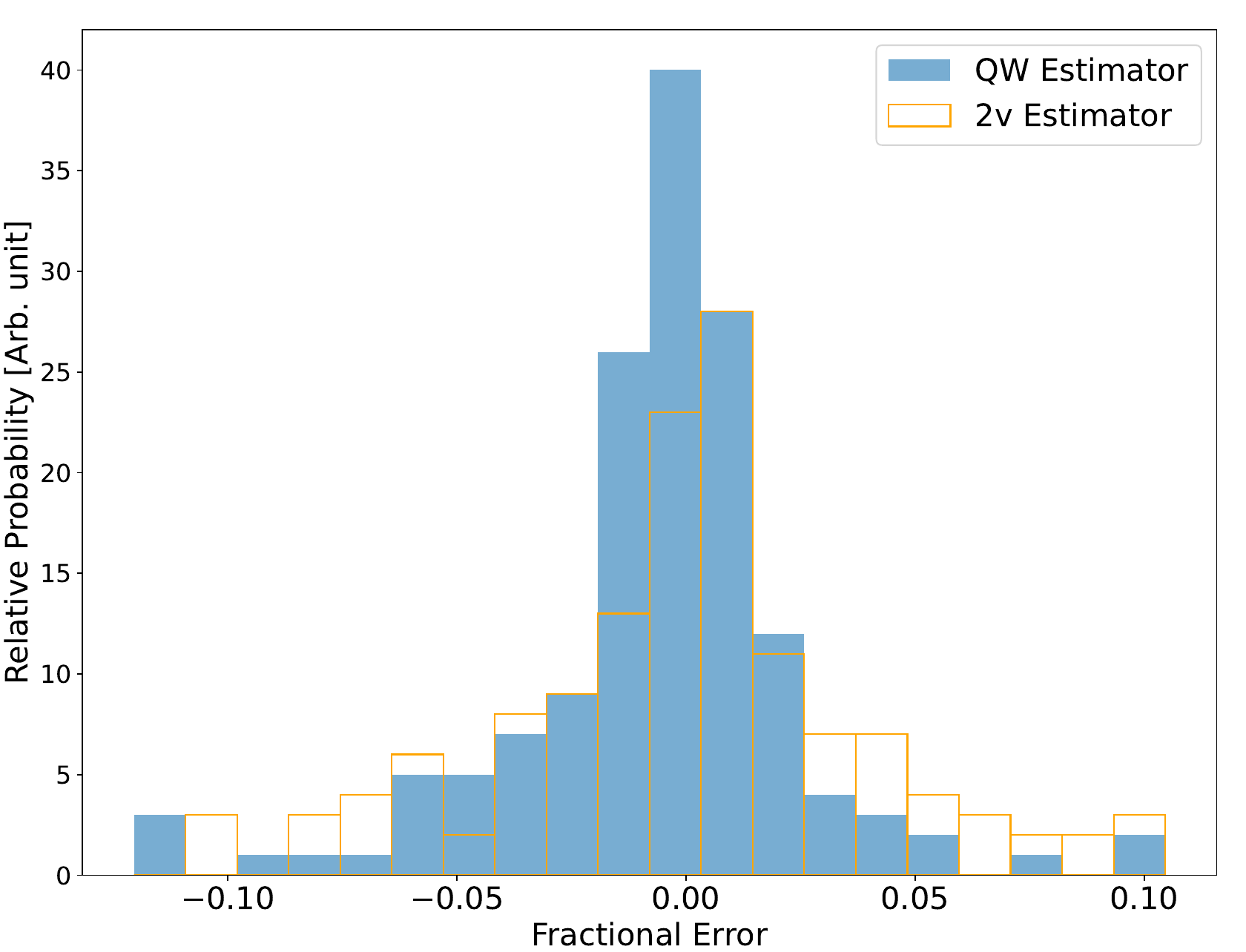}
    \caption{The error distribution of the \textit{2v estimator} and the \textit{Quad Jacobian estimator} (abbreviated as QW) after marginalising over all the 9 simulatedd configurations. Indeed, the QW estimators out-performed 2v estimator in the sense of having more tightly centred error distribution. In either case, the typical error is well within $\pm 5\%$. }
    \label{fig: estimators-performance-histogram}
\end{figure}
We wish to understand not only the performance of the magnification estimators, but also the reasons and conditions under which the estimators fail. 
To achieve this, we inject a fiducial lens system, which is characterised by an elliptical NFW profile. 
The ellipticity is configured so that it gives rise to a significant cusp in the caustics. 
We then place an extended source in various locations 
relative to the caustics, so that the resulting multiply-lensed images have different magnification ratios. 
This is to verify whether our estimators apply to more generic cases expected in the observational data, in which the lens arcs might be magnified unequally. 
(Lens arcs are usually equally magnified if they are close to a fold caustic, 
which has a larger lensing cross-section than other caustics with more complicated structures.)

We ran nine sets of simulations, populated on a $3\times 3$ configuration grid defined by the relative magnification across the arcs and the number of point sources in the source plane. 
In Fig~\ref{fig: estimators-performance-grid}, we show the fractional error of the magnification estimators as a function of the injected magnification ratios across the 9 simulations. 
Both estimators are less reliable when the number of point sources available on each of the lens arcs is limited, as demonstrated in the plots on the leftmost column. 
By increasing the number of point sources, the linear approximation on which our estimators are built
(see Section~\ref{sect: estimators}) 
becomes more accurate, as the typical separation between the point sources on the arcs is shorter. 
Consequently, the accuracy of the estimators gradually  improves from the middle column to the rightmost column, corresponding to a count of 30 point sources on each of the lens arcs.

Most of the multiply-imaged lens arc pairs possess a similar magnification, as expected from the image formation behaviour across a fold caustic. 
To understand more generic configurations, we also examined cases in which the lens arcs experience different 
magnifications. 
This physically corresponds to the rarer configurations in which the source galaxy is located near a higher-order caustic such as a cusp. 
Going from the first row of Fig~\ref{fig: estimators-performance-grid} to the bottom row, we skew the magnification ratio across the arcs in the simulation, from $-1$ (equal magnification) down to $-0.3$ (unequal magnification). 
\footnote{Here, the signed magnification ratio is always negative. This is a consequence of the fact that lensed image pairs are generated by crossing a specific caustic.} 
We find that the fractional estimation error of both our proposed estimators does not scale significantly with the injected magnification ratio. 
This is encouraging, as it shows that our estimators are generalisable to more complicated and realistic lens systems that feature unequal magnification across the arcs.

Comparing the performance of the \textit{QW estimator} and the \textit{2v estimator}, we find that the \textit{2v estimator} produces more biased estimates of the simulated magnification ratio, and 
this bias does not scale significantly with the specific magnification ratio of the particular point source under consideration. 
By contrast, while the \textit{QW estimator} is a comparatively more accurate estimator, it tends to produce smoother estimates, thereby underestimating the magnification ratio when the ratio is nearly equal, while overestimating it when the ratio is more disparate.

Finally, we marginalise over all simulated configurations 
to derive the error distributions for the two estimators. 
This is shown in Fig~\ref{fig: estimators-performance-histogram}. 
For both estimators, the fractional error is well within $\pm 5\%$. 
However, the \textit{2v estimator} exhibits a wider spread and heavier tails compared to the \textit{QW estimator}. 
Consequently, we decide to use the \textit{QW estimator} for our final analysis of the observed data.

\section{Identification of point sources on JWST image} \label{app: point-source-identification}
\begin{figure}
\centering
    \includegraphics[width=.45\textwidth]{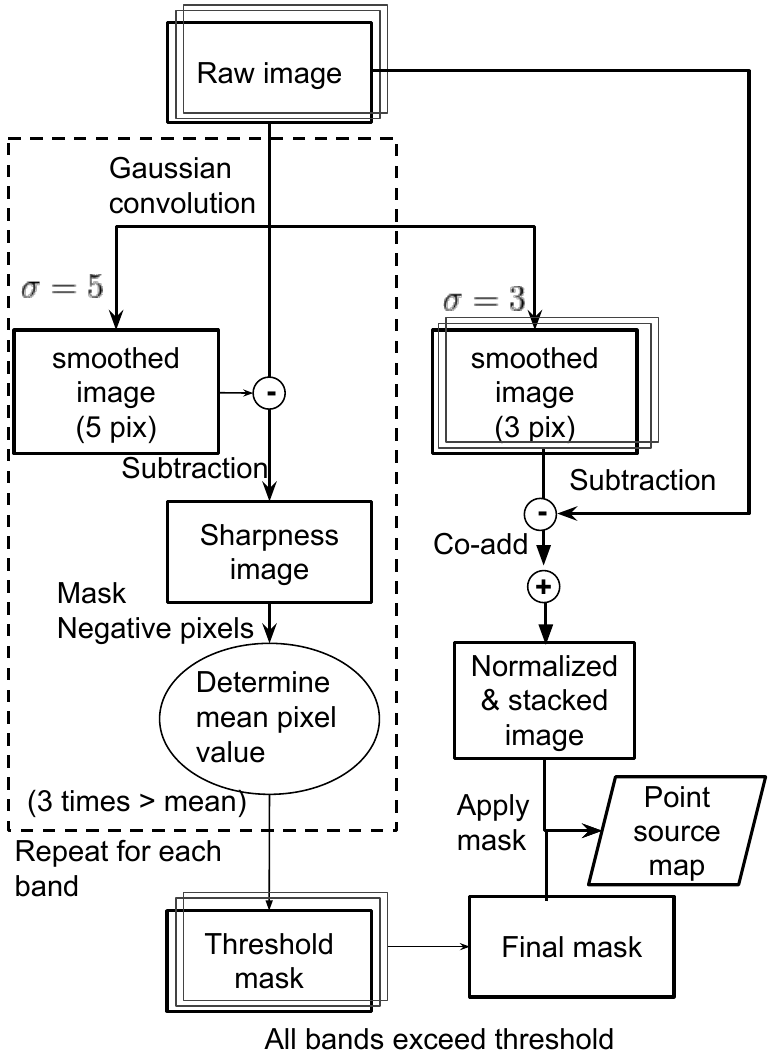}    
    \caption{The flow chart of the point source map generation process from the raw image data. } 
    \label{fig: point-source-filters-flowchart}
\end{figure}

The extraction of point sources along the lens arc is a non-trivial task owing to the brightness of the host lens arc. 
In this section, we outline a practical approach for the systematic identification of the point sources. 
This identification scheme can be further optimised. 
However, as the incomplete identification of available point sources does not bias our results, we adopt the following simple approach for proof-of-concept purposes.

The point sources extraction is based on a sequential image-filtering approach, 
as summarised in Fig~\ref{fig: point-source-filters-flowchart}. 
The images of each lens arc 
are separately smoothed by a Gaussian kernel with $\sigma = 5$ pixels. 
The raw image is then subtracted from this smoothed image, keeping only the pixels with positive values, 
resulting in a first filtered image that cleans most of the extended features. 
We then calculate the mean of this first filtered image and flag all pixels that are 3 times higher than the mean. 
This process is performed separately for the passbands F150W and F200W, resulting in 2 lists of flagged pixels. 
To remove noise, we generate a composite mask that retains only the pixels flagged in both filters. 
This procedure results in a masked, filtered image, where
the brightest point sources possess significantly higher pixel values than the remainder. 
In the final stage, the masked, filtered image is normalised again by co-adding the filtered images in both filters and normalising by a smoothed image generated by convolving the raw image with another Gaussian with $\sigma = 3$ pixels. 
We apply a thresholding scheme to this normalised image by requiring the pixel value to be larger than $0.3$, thereby generating a map of point source candidates. 
The resulting point source map is shown in Fig~\ref{fig: pairing-map}. 
The hyperparameters adopted in the above pipeline are 
largely based on trial and error. 
Further optimisation is left to future work.

The pairing of the point source maps for the two lens arcs is performed as follows. 
Using the raw image of the lens arc, we use its morphology to roughly determine the rotation and parity flipping required to transform between the 2 images. 
Note that in this step, we do not perform any scaling or warping --- the main purpose of the transformation is for visual inspection. 
Afterwards, we apply the same transformation to the point source maps. 
The relative positions of the point source candidates on the maps inform the identification of the correspondence between
the point sources on both images.

\bibliographystyle{mnras}

\bibliography{ref-jwst,ref-lensing,ref-wavedm,ref-forKeith,ref-cdm,ref-umbilic-lens,ref-smacs0723,ref-subhalos,ref-lens-quasar}
\bsp	
\label{lastpage}
\end{document}